\documentclass[a4paper,twoside]{article}

\newcommand{\affil}[1]{$^{\rm #1}$}
\usepackage[authoryear]{natbib}
\bibpunct{(}{)}{;}{a}{}{,}
\usepackage{multirow,epsfig,epstopdf}
\newcommand{\be}{\begin{equation}}
\newcommand{\ee}{\end{equation}}
\newcommand{\bc}{\begin{center}}
\newcommand{\ec}{\end{center}}

\usepackage{graphicx}
\usepackage{epsf}
\usepackage{color}
\usepackage{rotating}
\usepackage{txfonts}
\usepackage{geometry}
\usepackage{pifont}
\usepackage{natbib}
\date{} %Please leave the date blank
\def\lsim{\mathrel{\lower0.6ex\hbox{$\buildrel {\textstyle <}
 \over {\scriptstyle \sim}$}}}
\def\gsim{\mathrel{\lower0.6ex\hbox{$\buildrel {\textstyle >}
 \over {\scriptstyle \sim}$}}}

\title{\large\bf\flushleft Cold versus Warm Dark Matter simulations of a galaxy group}
\author{\parbox{\textwidth}{\flushleft
\vspace{-0.5cm}
{\it Noam I. Libeskind\affil{A,\dagger}, Arianna Di Cintio\affil{A.B,C}, Alexander Knebe\affil{B}, Gustavo Yepes\affil{B}, Stefan Gottl\"{o}ber\affil{A},  Matthias Steinmetz\affil{A}, Yehuda Hoffman\affil{D}, Luis A. Martinez-Vaquero\affil{E}}\\ %,, Arman Khalatyan\affil{A} and Kristin Riebe\affil{A}}\\
\vspace{0.4cm}
{\small \affil{A}\,Leibniz-Institut f\"{u}r Astrophysik, Potsdam, An der Sternwarte 16, 14482 Potsdam, Germany}\\
{\small \affil{B}\,Departamento de F\'isica Te\'orica, Grupo de Astrof\'{\i}sica,  Universidad Aut\'onoma de Madrid, Madrid E-28049, Spain}\\
{\small \affil{C}\,Physics Department "G. Marconi", Universita' di Roma "Sapienza", Ple Aldo Moro 2, 00185 Rome, Italy}\\
{\small \affil{D}\,Racah Institute of Physics, Hebrew University,  Jerusalem 91904, Israel}\\
{\small \affil{E}\,Grupo Interdisciplinar de Sistemas Complejos (GISC), Departamento de
Matem\'{a}ticas, Universidad Carlos III de Madrid, Leganes, Madrid, Spain}\\
{\small \affil{\dagger}\,Email: nlibeskind@aip.de}}}

\begin{document}
\twocolumn[
\begin{changemargin}{.8cm}{.5cm}
\begin{minipage}{.9\textwidth}
\vspace{-1cm}
\maketitle
%
%
%%%%%%%%%%%%%     ABSTRACT    %%%%%%%%%%%%%
%Abstract of no more than 200 words here.
\small{\bf Abstract:}

The differences between cold (CDM) and warm (WDM) dark matter in the formation
of a group of galaxies is examined by running two identical simulations where
in the WDM case the initial power spectrum has been altered to mimic a 1keV
dark matter particle. The CDM initial conditions were constrained to reproduce
at $z=0$ the correct local environment within which a ``Local Group'' (LG) of
galaxies may form. Two significant differences between the two simulations are
found. While in the CDM case a group of galaxies that resembles the real LG
forms, the WDM run fails to reproduce a viable LG, instead forming a
diffuse group which is still expanding at $z=0$. This is surprising since, due
to the suppression of small scale power in its power spectrum, WDM is naively
expected to only affect the collapse of small haloes and not necessarily the
dynamics on a scale of a group of galaxies. Furthermore the concentration of
baryons in halo center's is greater in CDM than in WDM and the properties of
the disks differ.\\

%%%%%%%%%%%%%     KEYWORDS    %%%%%%%%%%%%%

%%%%%%%%DO NOT EDIT%%%%%%%%%%%%
\medskip
\medskip
\end{minipage}
\end{changemargin}
]
\small
%%%%%%%%EDIT FROM HERE%%%%%%%%%%%%

\section{Introduction}
\label{sec:intro}

The current paradigm of galaxy formation, known as Cold Dark Matter (CDM), holds that structures in the universe grow in a bottom-up hierarchical fashion \citep[e.g.][]{1978MNRAS.183..341W}. The universe's initial conditions are conceived as a smooth roughly homogenous expanse of gas and dark matter (DM). In CDM, small perturbations imprinted on the primordial density field grow via gravitational instabilities, and then merge with each other to create the complex structures (such as clusters, groups of galaxies, galactic haloes, filaments, sheets and voids) we observed today. 

\begin{figure*}
\includegraphics[width=35pc]{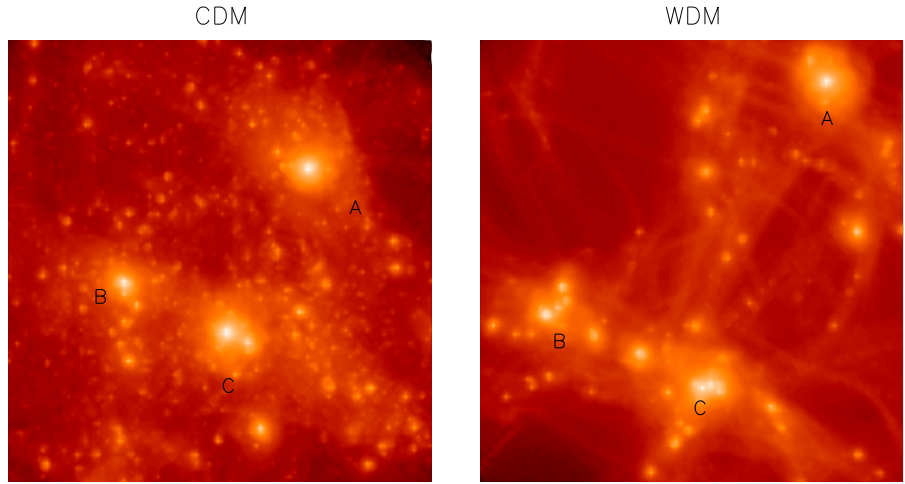}
\caption{A density map containing the three halos that make up the simulated group at $z=0$ in CDM (left) and WDM (right). The CDM group is more compact and collapsing while the WDM is more diffuse and still expanding. Each plot is projection of a 2$h^{-1}$Mpc cube.}
\label{fig:groups}
\end{figure*}

Warm DM (WDM), an alternative to CDM, suggests that initial perturbations below a certain mass cannot collapse and as such the smallest structures to form out of gravitational instability are fairly large \citep[e.g. $\sim 10^{10}h^{-1}M_{\odot}$][]{bode2001,zavala2009}. This is because the temperature of the DM particle at decoupling (specifically, whether it was relativistic or not) can cause the DM particle to escape from and erase the underlying density fluctuation. This process, known as ``free streaming'', inhibits the formation of small structures by gravitational collapse.

The initial power spectrum of fluctuations, which can be measured directly from the CMB, describes the degree of ``contrast'' in the density field and can be compared with the large scale clustering of galaxies observed in sky surveys (such as the SDSS or 2DF). 
These measurements probe the power spectrum on scales much greater than those
scales where the nature of the dark matter can be probed.

A number of suggestions as to the mass of DM particles have recently been proposed \citep[e.g.][]{2009ARNPS..59..191B,2009PhRvL.102t1304B} which corresponds to the lack of DM haloes less than $\sim 10^{6}M_{\odot}$ -- roughly the mass of the smallest DM - dominated dwarf galaxy. Indeed invoking a warmer flavor of DM \citep[such as a 2 keV sterile neutrino, see][]{Lovell11} may solve a number of issues related to dwarf satellite galaxies including the  ``Missing satellite problem" \citep{1999ApJ...524L..19M,1999ApJ...522...82K} as well as the ``Massive failure problem''  \citep{Boylan11,Boylan12}. Despite the many successes of CDM, there is thus more than just a hint that WDM may solve some of the fundamental problems in galaxy formation.

Regardless of the nature of the DM, the gravitational collapse of structures in the Universe is a highly non-linear process and can only be modeled by using numerical methods, such as $N$-body simulations \citep{2005Natur.435..629S} of the cosmic density field. Numerical simulations have successfully probed a myriad of scales: from the largest conceivable simulations of the universe \citep[e.g. the Horizon, Millenium-XXL and MultiDark runs][]{2011JKAS...44..217K,2012MNRAS.426.2046A,2011arXiv1109.0003R},  through clusters \citep[e.g. the Phoenix project][]{2012MNRAS.425.2169G}, to Milky Way (MW) type galaxies filled with small substructures \citep{2008MNRAS.391.1685S,2009MNRAS.398L..21S}. 

Within the CLUES project\footnote{http://www.clues-project.org}  we have used
constrained simulations to shown that the specific environment of the Local
Group is an important ingredient in the formation of the Milky Way and
Andromeda galaxies
\citep[e.g.][]{2005MNRAS.363..146L,2010MNRAS.405.1119K,Libeskindetal2011,2011MNRAS.411.1525L,2011MNRAS.412..529K}. Indeed
the often used term ``MW-type galaxy'' which lumps all galaxies in haloes of
$\sim10^{12}M_{\odot}$ together, may  be considered a stereotype given the
wide differences in merger history, morphology, and other properties among
these galaxies
\citep[e.g.][]{2009MNRAS.395..210D,2011ApJ...743..117B,2011arXiv1107.0017F}. 
Since the simulations  can be directly compared with observations  constrained
simulations are extremely useful to study the formation of the Local Group
galaxies
\citep[e.g.][]{2011arXiv1107.2944K,DiCintio11,DiCintio12,DiCintio12b,2012MNRAS.419L...9D}.

Constrained simulations have also been used to study the velocity function of dic galaxies in the Local
Volume by \cite{zavala2009}. By using a simple model to populate halos with disk galaxies, \cite{zavala2009} showed  that the velocity functions in the two regions explored by the ALFALFA survey agree quite well both CDM and WDM cosmologies, as long
as one considers massive galaxies with circular velocities in the range in the range between $80km s^{−1}$
and $300 km s^{−1}$. However,  for galaxies with circular velocities below $80
km s^{-1}$  only the predictions of a 1keV WDM particle, agrees with observations. On the other hand, 
 at a circular velocity of $\approx 35 km s^{-1} $ the CDM scenario predicts about 10 times more sources than
observed. 

 Using the same set of simulations as \cite{zavala2009},  \cite{2009MNRAS.399.1611T} found that the observed spectrum of mini-voids in the local volume is in good agreement with the WDM model but can hardly be explained within the CDM scenario.

Given the importance of the Local Group on the formation of the MW, in this
paper we examine the effect of the type of DM assumed, on forming such a group. We use the same model as \cite{zavala2009} but
run gasdynamical simulations with much higher resolution as described in Section 2.In Section 3 we study the cosmography of the
 simulated groups and in Section 4 the internal halo properties. In Section 4
 we summarize and discuss our results.

\section{Simulations}

In this section we describe briefly the numerical methods used to run our
simulations as well as the methods to identify halos in the simulation. We
refer the reader to \cite{2010MNRAS.401.1889L} for details.  As mentioned
before the original CDM simulation was constrained by present day observations
of our local universe
\citep{1997ApJS..109..333W,2001ApJ...546..681T,2004AJ....127.2031K,2002ApJ...567..716R}. Initial
conditions are then produced following the method described by
\citep{1991ApJ...380L...5H}. The zoomed DM initial conditions for a $2h^{-1}$
Mpc sphere were generated following the prescription set out in
\cite{2001ApJ...554..903K}. The reader should note that the constraints we have applied to the initial conditions are on linear scales at $z=0$ and are identical in the two cosmologies. The unconstrained phases, namely the power responsible for the internal dynamics of the groups embedded in the constrained realizations are effectively random. ``Effectively'' because they have been selected in the CDM case (by trial and error)  to produce a group which resembles the LG in terms of number, mass, geometry and kinematics of three galaxies. Therefore, an unconstrained random realization which produced a LG looking candidate with CDM initial conditions would have equally sufficed for the purposes of our study.

Gas particles are included in the high resolution regions of both the WDM and
CDM initial conditions with a mass of $m_{\rm GAS} = 4.4 \times 10^{4} h^{-1}
M_{\odot} $: during the evolution of the simulation they may spawn star
particles (see below), whose mass is $m_{\rm STAR} = 0.5m_{\rm GAS} = 2.2
\times 10^{4} h^{-1} M_{\odot} $. Star, gas and high resolution DM particles are all softened on the same length scale of 150 $h^{-1}$pc. Star formation rules are described in detail
in \cite{2010MNRAS.401.1889L}. The \cite{2003MNRAS.339..289S} method is used
to model gas in the interstellar medium. A uniform but evolving ultra-violet
background is switched on at $z=6$ \citep{1996ApJ...461...20H}. Only atomic
cooling is assumed. Cold gas cloud formation by thermal instability, star
formation, the evaporation of gas clouds, and the heating of ambient gas by
supernova driven winds all occur at the same instant. Each star formation
event injects energy and metals into the ISM instantaneously. Feedback from SN
explosions is modeled kinetically using the stochastic approach developed by
\citet{2003MNRAS.339..289S}.

 The PMTree-SPH MPI code \textsc{Gadget2} \citep{2005MNRAS.364.1105S} is used
 in both runs to to simulate the evolution of a periodic cosmological box with
 side length of $L_{\rm box}=64 h^{-1} \rm Mpc$. Using the same sub-grid
 physics we modified only the initial power spectrum of
 fluctuations to simulate a WDM  model. Since the phases of the constrained
 initial conditions in both cases are identical, any differences in galaxy or
 halo properties is directly due to the effect of changing the DM power
 spectrum.  Both runs employ cosmologies that assume WMAP3 parameters \citep{2007ApJS..170..377S},
i.e.  $\Omega_m = 0.24$, $\Omega_{b} = 0.042$, $\Omega_{\Lambda} = 0.76$. The rms mass fluctuation in spheres of 8~Mpc is   $\sigma_8 = 0.73$ and $n=0.95$ is the slope of the power spectrum. 

\begin{figure}
\includegraphics[width=15pc]{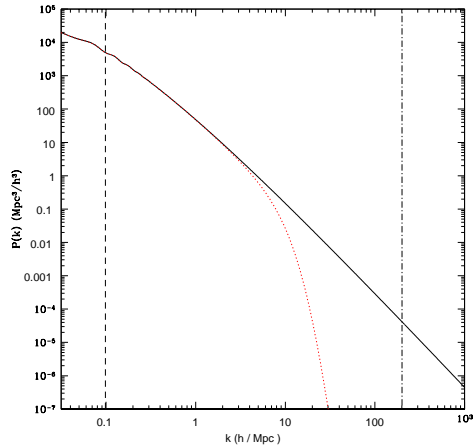}
\caption{The power spectrum used in this work. In black we show the CDM power spectrum, in red, the WDM power spectrum. The vertical dashed lines indicate the $k$ interval used to generate the initial conditions, from the fundamental mode ($k\sim2\pi/L_{\rm box}\approx0.1$)  to the Nyquist frequency ($k\sim 200$).}
\label{fig:powspec}
\end{figure}

When simulating WDM we suppress the power spectrum below scales representative of a 1keV WDM particle (see Fig.~\ref{fig:powspec}). The initial conditions are generated by rescaling the CDM power spectrum and fitting it with an approximation to the transfer function representative of the free streaming effect of WDM particles \citep{2005PhRvD..71f3534V}. The the free-streaming length of such a WDM particle is $350h^{-1}kpc$, which corresponds to a filtering mass of $\sim1.1x10^{10}h^{-1}M_{\odot}$ \citep{bode2001}: the WDM power spectrum, shown in Fig.~\ref{fig:powspec} thus contains a sharp cut-off at this free-streaming length.

\label{sec:sims}

\label{sec:sims}
\begin{figure*}
\includegraphics[width=35pc]{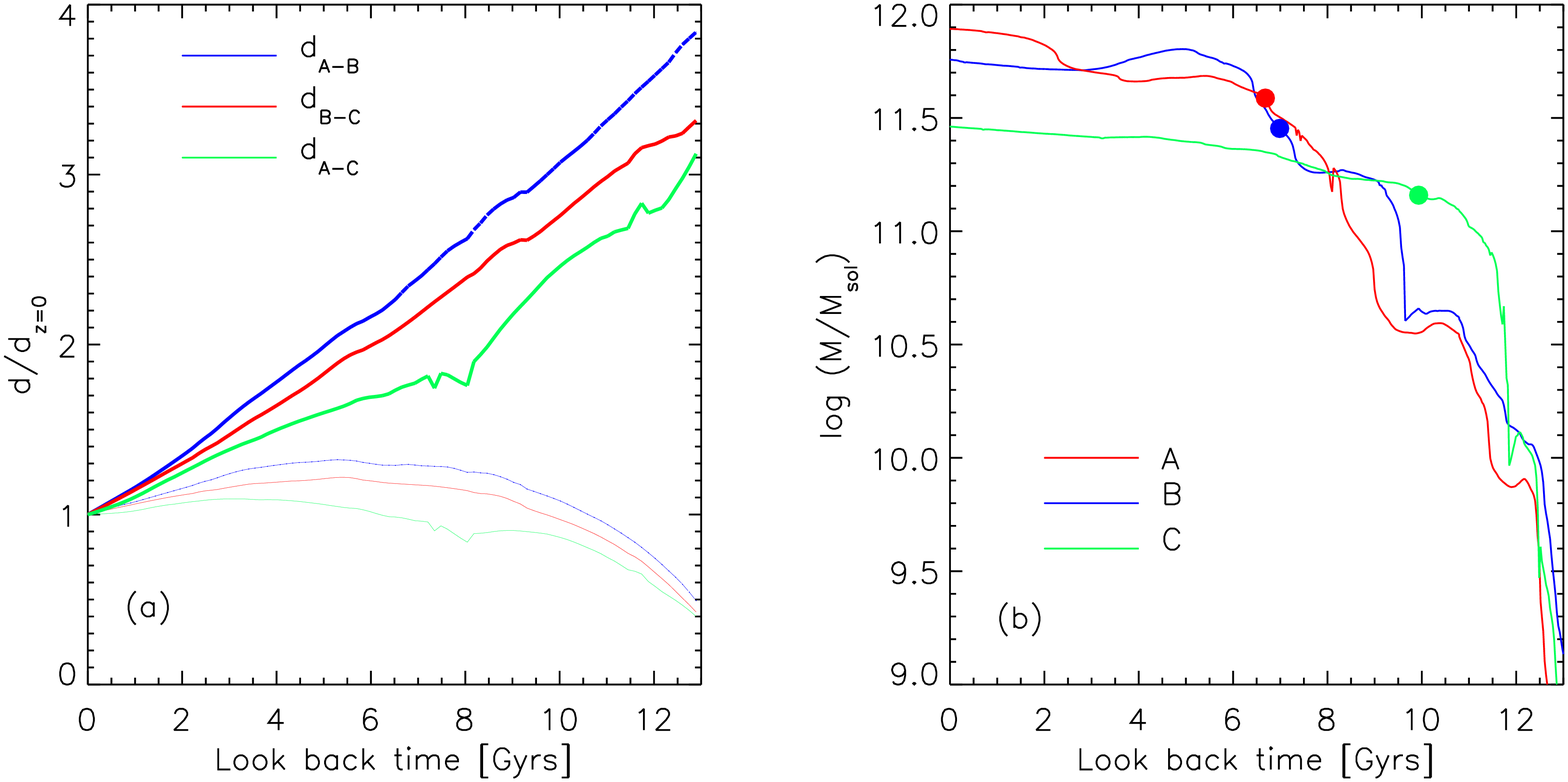}
\hspace{1.5 cm}\includegraphics[width=17pc]{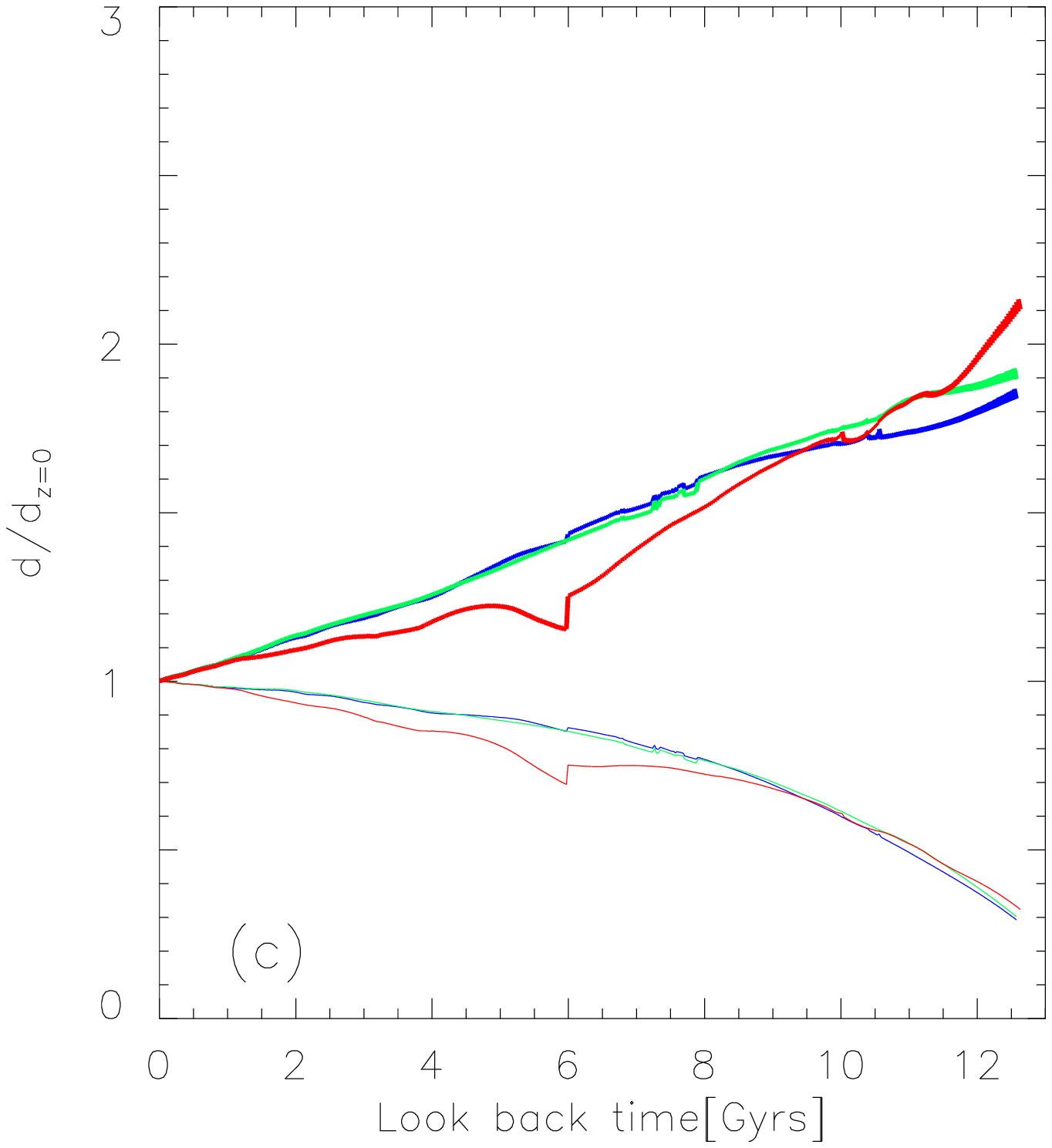}
\hspace{0.3 cm}\includegraphics[width=17pc]{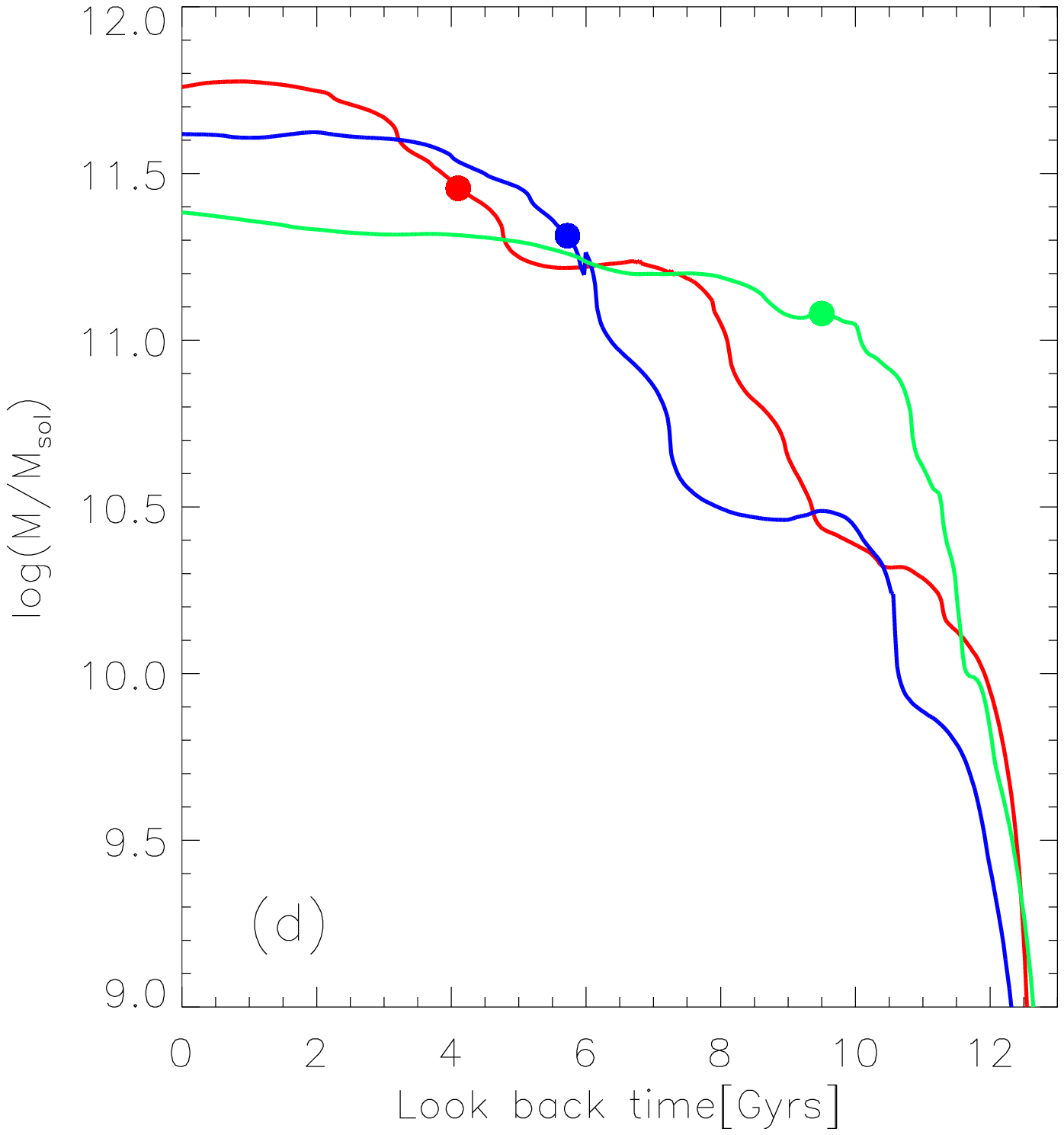}
\caption{{\it Upper Panels:} CDM; {\it Lower Panels:} WDM. \textit{Left Panels (a,c):} The physical (\textit{thin line}) and co-moving (\textit{thick line}) distance as a function of look back time between the three pairs of LG haloes. We show the distances between the A and B in blue, the B and C in red and A and C in green. Each curve is normalized to its $z=0$ value which can be found in Table~1. \textit{Right Panels (b,d):} The mass growth for halo A (red),  B (blue), and C (green) as a function of look-back time. The solid dots denote the time at which half the $z=0$ mass was assembled.}
\label{fig:evol}
\end{figure*}

\begin{figure*}
\bc
\includegraphics[width=35pc]{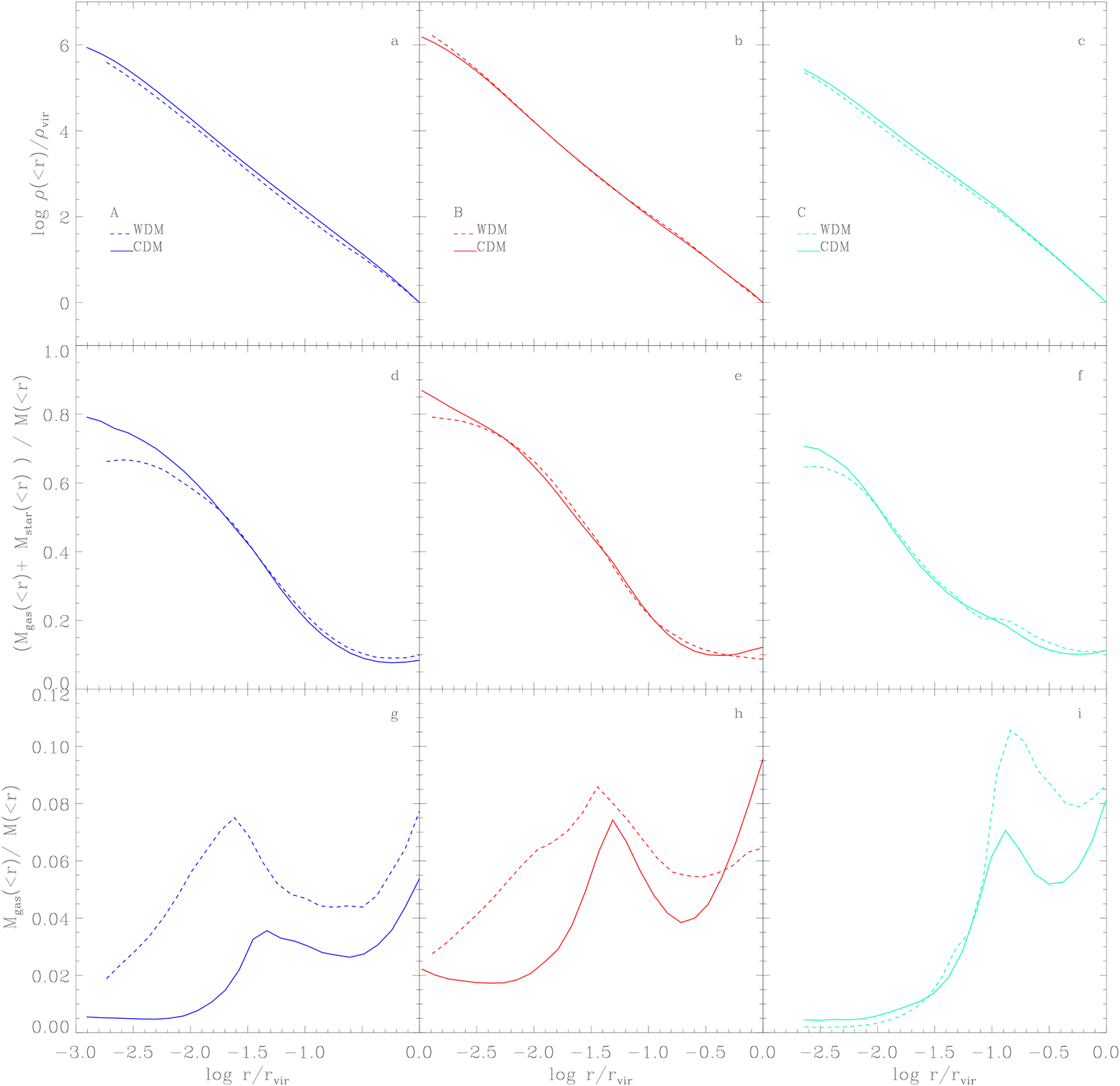}
\ec
 \caption{Internal properties of the three main haloes simulated as function of radius. Properties for halo A (red, left panel), B, (blue, center panel) and C (cyan, right panel) are shown for WDM (dashed) and CDM (solid). \textit{Top row:} Density profile. \textit{Middle row:} Baryon fraction. \textit{Bottom row:} Gas fraction.}
\label{fig:densprof}
\end{figure*}

In order to identify halos and subhaloes in our simulation we have run the
MPI+OpenMP hybrid halo finder \texttt{AHF}\footnote{Publicly available ar \texttt{http://popia.ft.uam.es/AHF}}. We refer the reader to the code description papers \citep{GIll04b,2009ApJS..182..608K} for details. \texttt{AHF} locates local over-densities in an adaptively smoothed density field as prospective halo
centers. The potential minimum of each density peak is then calculated; bound particles are then associated to possible haloes.

In the WDM simulation, discreteness effects which can cause haloes below a specific limit mass ($M_{lim}$) to arise from the unphysical numerical fragmentation of filaments, is an issue. In order to protect our analysis against these artificially formed haloes we use the value of $M_{lim}$ provided by \citet{Wang2007} as the minimum trusted mass for a halo in the WDM simulation. Their expression, originally based upon Hot Dark Matter models, reads $M_{lim}=10.1\bar{\rho}d/k_{peak}^{2}$, where $\bar{\rho}$ is the mean density, $d$ is the mean interparticle separation, and $k_{peak}$ is the wavenumber at which $\Delta^2(k) = k^3P(k)$ reaches its maximum. In our WDM run, where the power spectrum has been modified to correspond to a $1 keV$ particle, the values of this limiting mass is $M_{lim}\sim 2.6 x 10^7 M_{\odot}/h$, which corresponds roughly to a 100 particle limit. In practice in both the CDM or  WDM simulation, only objects whose mass is greater than 500 particles are used. We note that since the simulations have identical baryonic physics, particle mass and spatial resolutions any of the differences reported here are due entirely to the nature of the DM model.

\section{Cosmography}
\label{sec:cosmography}

We begin with a cosmographic description of the two simulated groups. Our simulations produce three dominant objects which we name galaxy A, B and C in decreasing mass. In the CDM case these closely resemble the Milky Way (MW), Andromeda (M31) and Triangulum (M33).  An image of the two groups can be seen in Fig.~\ref{fig:groups}. Two salient aspects of WDM are immediately apparent from this figure: (1) there are far fewer small substructures and (2) the two groups differ substantially, cosmographically speaking.

In Figure \ref{fig:evol}(a,c) we show the co-moving and physical distance between the three pairs of group members as a function of look back time, normalized to the $z=0$ value. In the CDM simulation, the physical separation of each pair of galaxies reaches a maximum ``turn-around'' (at a look back time of around 6 Gyrs for galaxy B-A and galaxy B - C pair and around a few Gyrs later for galaxy A - C).  In the WDM simulation this is not the case: the physical distance between each pair of haloes at every redshift is smaller than the corresponding distance at redshift zero, indicating that the Hubble expansion is the dominant force at every epoch and that all three pairs of galaxies have yet to begin approaching each other. Accordingly, the group is more compact in CDM than in WDM. Using these specific initial conditions,  over densities that turn around and are on a collision course at a given epoch in cosmic time in CDM, have yet to approach each other in WDM: where CDM produces an attracting, collapsing  group of galaxies, WDM produces a still expanding version. This is our first result: \textit{Using initial conditions, whose only difference is a suppression of small scale power, the defining dynamics of the a group of galaxies are completely different in CDM and WDM, with the former predicting an attracting group that resembles the LG, while the later predicting a still expanding one}.

The co-moving distances (shown as the thick lines in Fig.~\ref{fig:evol}a,c) show monotonic attractions. In the WDM case the simulated haloes are closer to each other (relative to their $z=0$ distances) at early times than the CDM halos. In the CDM case, by $z=0$ the haloes have been brought closer. Note that the small kinks in the A -C system (CDM) and B - C system (WDM case) appear due to false identification of the main progenitor in the merger tree construction at a given snapshot.

\begin{table*}
\begin{center}
 \begin{tabular}{l l l }
 \hline
 \hline
Property &  CDM  Group & WDM  Group \\
   \hline
   \hline
  
$M_{\rm A}$  & $7.49\times10^{11} M_{\odot}$ & $5.75\times10^{11} M_{\odot}$ \\
$M_{\rm B}$  & $5.48\times10^{11} M_{\odot}$ & $4.15\times10^{11} M_{\odot}$ \\
$M_{\rm C}$ & $2.78\times10^{11} M_{\odot}$ &$2.42\times10^{11} M_{\odot}$  \\
$d_{\rm A-B}$ & 1.22 Mpc & 2.26 Mpc \\
$d_{\rm A-C}$ & 1.37 Mpc & 2.34 Mpc \\
$d_{\rm B-C}$ & 0.79 Mpc & 1.22 Mpc \\
$V_{\rm A, B}$ & -110~kms$^{-1}$& 29~kms$^{-1}$  \\
$V_{\rm A, C}$ & -85~kms$^{-1}$&  35~kms$^{-1}$ \\
$V_{\rm B, C}$ & -4~kms$^{-1}$&  42~kms$^{-1}$ \\
    \hline
    \hline
 \end{tabular}
 \end{center}
\caption{The $z=0$ properties of a simulated group in CDM and  WDM. Note that the WDM group has little resemblance to the CDM one (which closely matched the real LG, see~\citet{2010MNRAS.401.1889L}. We show the following properties: the mass of halo A, B and C ($M_{\rm A}$, $M_{\rm B}$ and $M_{\rm C}$), the distance between halos A, B and C ($d_{\rm A, B}$, $d_{\rm A-C}$ and $d_{\rm B-C}$), and the relative line of sight velocity for each pair ($V_{\rm A, B}$, $V_{\rm A, C}$, and $V_{\rm AB, C}$} 
\label{table:LGprops}
\end{table*}

We now examine the evolution of the three individual group members by examining the mass accretion history shown in Fig.~\ref{fig:evol}(b,d). In both the CDM and WDM run, the two most massive galactic haloes  (A and B) show jumps in the mass accretion history characteristic of merger activity occurring more or less continuously. Often, these haloes appear to lose mass after a violent major merger. This is because of the unique merger history of these objects - violent mergers may bring material into the virial radius that is bound at one redshift, but which may become unbound and flung out at a later time. The smallest halo (C) on the other hand shows little evidence of major mergers in its past.

Although the mass growth histories look similar, in fact they differ slightly. The time at which half of the $z=0$ mass has been assembled is shown in each plot as a filled circle. In the WDM simulation, each halo assembles 50\% of its mass later with respect to the CDM model. Specifically, in the WDM run halo A, B and C accrete half-mass at a look-back time of $\sim4$,~$\sim6$, and $\sim9.5$Gyrs, respectively. In the CDM case this occurs at $\sim7$, $\sim7$ and $\sim 10$Gyrs: that is $\sim3$, $\sim1$, $\sim0.5$ Gyrs earlier. Since  B and C are smaller mass haloes, their half mass times are considerably earlier and the delay is considerably smaller than for halo A.

A characteristic feature of the WDM model emerges here: the finite primordial phase-space density due to the large thermal velocities of the particles causes most of the mass to undergo gravitational collapse at later redshift ($z<5$), resulting in the suppression of halo formation at higher redshift \citep{bode2001}. Halo collapse is thus delayed with respect to the CDM model. Although not a new result, this finding directly informs the main differences we find between CDM and WDM.

\section{Internal Halo Properties}
\label{sec:structure}

How do the different cosmographies and histories change the internal structure of each of our three LG objects? In Fig.~\ref{fig:densprof}(a)-(c) we show the density profile of the three LG members in both WDM (dashed) and CDM (solid) simulations. All density profiles are standard NFW fits, and in all three cases the WDM is nearly indistinguishable from the CDM. That said, owing to the lower mass of the WDM haloes,  their density profiles are systematically shifted to slightly lower densities.

In Fig~\ref{fig:densprof}(d)-(f) we show the cumulative baryon fraction as a function of radius. Again, WDM and CDM show broad similarities in shape and value of the baryon fraction. In the inner parts, WDM shows a systematically lower baryon fraction. At around $\sim0.03r_{\rm vir}$, the total fraction of internal mass in baryons is roughly the same in both cosmologies. Towards the outer parts of the halo, the baryon fraction of both cosmologies drops, reaching the cosmic mean of $\sim0.1$ at the virial radius. 
That CDM haloes have more concentrated baryons is likely due to a number of combining factors: their earlier formation time, their greater mass and thus their deeper potential. This is our second main result: \textit{WDM haloes have lower baryon fractions in their inner parts where baryons dominate, than CDM haloes}.

The baryonic properties of the three Local Group members are summarized in Table~2.

The fraction of mass in a gaseous component is presented in Fig.~\ref{fog:densprof}(g)-(i). Although each halo shows different specific behavior, some interesting similarities exist. Firstly, the fraction of mass in gas is almost always greater in WDM than in CDM. This is true for all radii in halo A, and for radii greater than $0.03r_{\rm vir}$ for halo B and C (although in halo B, there is more gas in CDM for $r<0.2r_{\rm vir}$). The  higher gas fractions in WDM may inhibit infalling substructures  from depositing their material in the center of the halo thereby suppressing the baryon fraction in the inner parts of WDM halos, as seen in Fig.~\ref{fig:densprof}(d)-(f).

\begin{table*} 
  \begin{center}
    \begin{tabular}{llllllll}
\hline
\hline
             Galaxy  & Property        & & CDM &&&WDM       \\
\hline
  & &  TOTAL        &  GAS        &  STARS  &  TOTAL        &  GAS        &  STARS \\
\hline
\hline
\multirow{9}{*}  & $N_{\rm vir}$       ($10^{6}$)    &  4.2          &  1.3        &  0.65  & 2.9 &   0.66   & 0.43   \\      
                   A   & $M_{\rm vir}$ ($10^{11} M_{\odot}$)     &  5.5          &  0.52       &  0.14    & 4.2  &  0.27 &  0.094   \\
                      & $f_{\rm b,\rm vir}$                         &  0.12 && &  0.09     \\
		  
\hline
\multirow{9}{*}   & $N_{\rm vir}$       ($10^{6}$)    &  2.9          &  0.53       &  0.55  & 2.2    & 0.56    &   0.30\\      
                 B     & $M_{\rm vir}$ ($10^{11} M_{\odot}$)     &  4.0          &  0.21       &  0.12  & 3.0 &  0.23 & 0.066     \\
                      & $f_{\rm b, \rm vir}$                         &  0.08 && & 0.09      \\
                       \hline
\multirow{9}{*}  & $N_{\rm vir}$       ($10^{6}$)    &  1.5          &  0.40       &  0.29     & 1.3   & 0.36 &  0.19 \\      
              C        & $M_{\rm vir}$ ($10^{11} M_{\odot}$)     &  2.0          &  0.17       &  0.064   &   1.8 & 0.15 & 0.040  \\
                      & $f_{\rm b,\rm vir}$                         &  0.11 &&  & 0.11      \\
                   \hline
                \hline
    \end{tabular}   
 \caption{Properties of the three main galaxies in the CDM and WDM simulation. For each halo we show the number ($N_{\rm vir}$) and mass ($M_{\rm vir}$) of stars, gas and all particles within the virial radius. We present the baryon fraction within the virial radius ($f_{\rm b, \rm vir}$).}
   \label{tab_LGhrG}
  \end{center}      
\end {table*}

%%%%%%%%%%%%%%%%%%%%%%%%%%

Both gas and stars form well defined discs, a consequence of the star formation rules we have used. This can be quantified by performing a dynamical bulge-disc decomposition. There are a number of ways this is done in the literature \citep[e.g][]{Abadi03b,Scannapieco10,Sales12}. In this work we dynamically decompose star and gas particles within the inner 10~kpc into disc-like and bulge-like components using two methods, one for each component \citep[as in][]{2013MNRAS.428.2039K}. For both methods a ``disc-axis'', taken to be the total angular momentum of  all baryonic particles within 10kpc, must be assumed.

For gas particles we follow \cite{Scannapieco10}; the component of each particle's angular momentum in this direction ($J_{z}$) is computed and compared with the angular momentum a particle would have at that radius if it were on a circular orbit. The ratio $J_{z}/J_{\rm circ}$ is computed where:
\begin{eqnarray}
	J_{\rm circ}&=& r  \times v_{\rm circ}\\
		&=&r \times \sqrt{\frac{GM(r)}{r}}
\end{eqnarray}
Here $M(r)$ is the total mass (including DM) within a radius $r$. Note that in this formulation, particles with $J_{z}/J_{\rm circ}\approx 1$ are on circular orbits and thus compose a disc. Note that $J_{z} > J_{\rm circ}$ and thus the ratio ranges from (0,$\infty$).

For star particles we follow \cite{Abadi03b} and compare the component of the angular momentum in the $z$-direction with  the angular momentum of a circular orbit of the same energy, $J_{\rm c}(E)$. First, the total (kinetic plus potential) energy of each particle is computed. Since circular orbits maximize angular momentum, the maximum value of $J_{z}$ for all particles with a given energy is taken as $J_{\rm c}(E)$. In this case the ratio $J_{z}/J_{\rm c}(E)$ is confined to the interval [-1,1], where negative values imply counter-rotation with respect to total angular momentum of all baryonic particles within 10kpc.

Two different methods for gas and star particles are used because of the nature of the the methods themselves. The Abadi et al. method is more appropriate for $N$-body particles where the energy is simply kinetic plus potential. Gas particles have an extra component (internal energy) which informs their dynamics. In this case its better to use the Scannapieco approach.

\begin{figure*}
\begin{center}
\includegraphics[width=30pc]{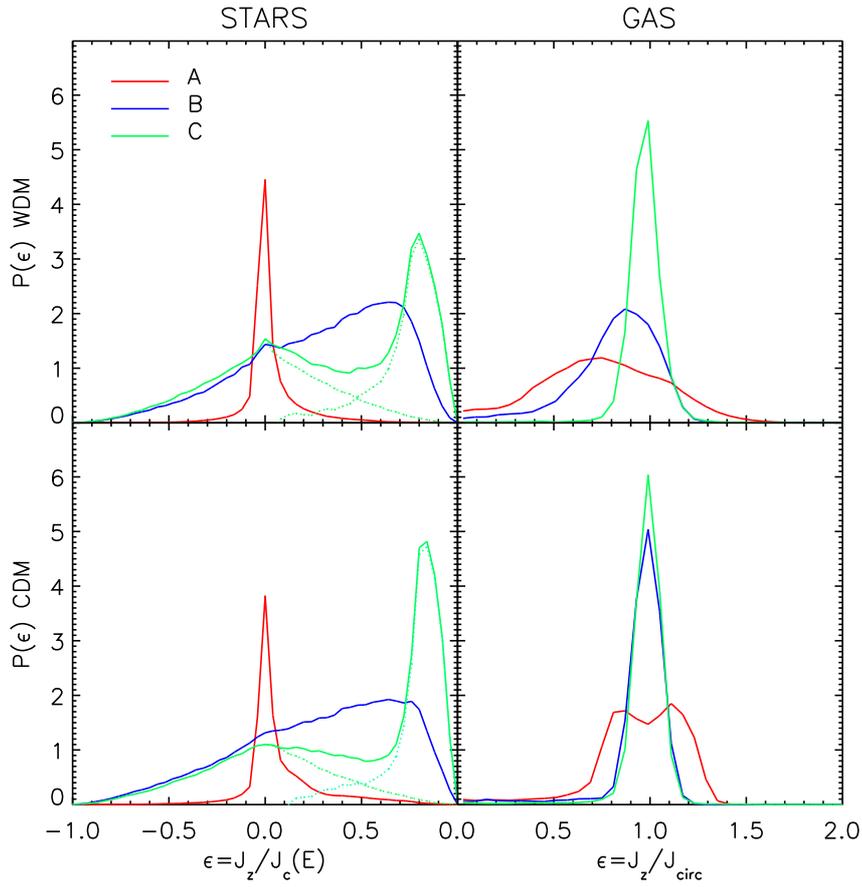}
\caption{The fraction of stellar (gas) particles within 10kpc at $z=0$ with a given ratio of $J_{z}/J_c(E)$ ($J_{z}/J_{\rm circ}$) for the galaxies in halo A (red), B (blue) and C (green). Particles with $J_{z}/J_{\rm circ}\approx 1$ are on circular orbits and thus compose a disc. Note that the gas particles nearly all constitute a disc, while star particles populate both disc and bulge components. The dip at $J_{z}/J_{\rm circ}\approx 1$ of the gas component of galaxy A is due to warping of the disc. The dotted green line indicates a decomposition into bulge and disc star particles for galaxy C.}
\label{fig:bdecomp}
\end{center}
\end{figure*}

In Fig.~\ref{fig:bdecomp} we present histograms of  $J_{z}/J_{\rm c}(E)$ (left column, star particles) and  $J_{z}/J_{\rm circ}$ (right column gas particles) for the CDM (bottom row) and WDM (top row) simulations. In the CDM simulation, gas in both the B and C clearly define a very thin disc, while A's gas is less ordered. Star particles on the other hand show a well defined disc in C's case, a ``fat'' disc in B's case and no disc in A's case

In the WDM run, the gas particles of halo C appear to define a clear disc while halos A and B have poorer gaseous discs. With respect to the stars we see a similar situation to the CDM case. Halo C has a disc component, B has a thicker disc and A has no real disc.

Due to the fact that halo A has a significative stellar bulge, the corresponding star particle histogram has been rescaled by a factor of four with respect to the stellar particle histogram of the other two galaxies, for both the CDM and WDM runs (the peak of the star component of halo A was $~20$ in both runs).

It is interesting to note that the discs of B and C are smaller in the WDM case than in the CDM case. This may again be a result of the delayed formation time of WDM haloes and the consequent lower mass. It is interesting that the bulge component (namely the peak at  $J_{z}/J_{\rm c}(E)=0$) seems to be roughly of the same size in both A and B.

Note that the dip at $J_{z}/J_{\rm circ}\approx 1$ in the gas distribution of A in the CDM simulation is due to a warping of the disc. 

C is the only galaxy that, owing to its quiet merger history, forms a clearly identifiable stellar disc, decomposed in Fig.~\ref{fig:bdecomp} into bulge and disc components (see dashed lines). The total mass in each component is similar: In CDM, 44\% and 56\% of halo C's galaxy is attributed to a bulge and disc, respectively. These fractions are nearly perfectly inverted in WDM: 45\% and 55\% of  halo C's stellar component are disc and bulge, respectively.

Although our sample size is small, we note that one of the more unanticipated consequences of haloes forming later in WDM, is their smaller and thicker disc. Indeed this may simply be a reflection of the different dynamical environments of the two groups. More work on the relationship of disc thickness to DM particle mass is encouraged to see if one can constrain the other.

Since the dynamical decomposition indicates that the galaxies within each halo differ substantially, it is perhaps no surprise that so too do their star formation histories. Although not shown here, the SFR (being a reflection of the merger history) is quantitatively very different in the two cosmologies.

As expected (and seen elsewhere) our WDM simulation has far fewer satellites than our CDM simulation.  WDM produces roughly the same number ($\sim 20$) of subhaloes as satellites observed to be in orbit about the Milky Way. However it is unclear if, owing to feedback and other star formation suppression mechanisms, WDM subhaloes are luminous enough to match the MW's satellite luminosity function.

\section{Summary and Discussion}
\label{sec:conclusion}
Since the temperature of the DM particle at decoupling determines its ability to ``free-stream'' out of potential wells, it also sets the scale at which structures are able collapse. In principle this characteristic can be used to constrain DM to be either ``cold'', ``warm'', or ``hot''. Hot DM, such as Neutrinos which travel at relativistic speeds, were at first hailed as the solution to the DM problem but have now been effectively ruled out \citep{2005PhR...405..279B} since they can escape most potential wells and prevent structures from formation via gravitational instability. Cold DM (CDM), on the other hand moves non-relativistically and as such is able to collapse into objects as small as an Earth mass \citep{2010ApJ...723L.195I}. The prediction of small substructures embedded in larger objects is a generic feature of the CDM model and, since such objects are unobserved in the Milky Way, this has lead to the famed ``Missing satellite problem'' \citep{1999ApJ...524L..19M,1999ApJ...522...82K}, often dubbed a crisis for CDM. Astrophysical process \citep[such as photo-evaporation of star forming gas due to UV radiation, see][]{2003ApJ...599...38B} are invoked to inhibit the gas cooling into small subhaloes. These process do not erase substructures, they simply ensure that they remain non-luminous. A large population of dark subhaloes  detectable via gamma ray emission from DM annihilation \citep{2003MNRAS.345.1313S} or via  strong gravitational lensing of background sources \citep{2009MNRAS.398.1235X}, is thus predicted albeit unobserved, in the Milky Way halo.

As a result of the apparent failures of CDM in over producing and HDM in underproducing the number of dwarf satellites around the Milky Way, warm DM (WDM), has recently been suggested and studied \citep[by e.g.][among others]{bode2001,2001ApJ...559..516A,2002MNRAS.329..813K,2008MNRAS.386.1029K,2010MNRAS.404L..16M,Lovell11,2013MNRAS.428..882M}. In this paper, we have used a set of initial conditions that constrain scales unaffected by the nature of the DM to test the effect of the type of DM on a group of galaxies (i.e. $\sim 1$Mpc). Within the scales that are still linear at $z=0$ (the ``local environment'') a group of galaxies that in CDM resembles the local group (LG) is resimulated at high resolution, with gasdynamics. In the CDM run, this local group includes three galaxies that  have the same mass, geometry and kinematics as the MW, M31 and M33. Thus our simulation allows us to study in detail the merger history and internal structure of these galaxies as well as their baryonic properties. Since the local environment has been kept identical, we can directly measure the effect the type of DM has on our CDM LG.

The main difference between our CDM and WDM simulations is that structure formation is delayed in WDM. This is a direct result of the suppression of small scale power which, owing to the lack of mergers below the filtering mass,  means that it takes longer for haloes to grow to a given mass. The greatest effect this has is to inhibit the collapse of a group of galaxies in WDM. All our results regarding the differences in the galaxies themselves, follow directly from this main difference.

\begin{itemize}
\item A group of galaxies which at $z=0$ closely resembles the LG in CDM, is dynamically very different in WDM. Whereas in CDM the group is collapsing and is compact, in WDM it is still expanding and is much more diffuse.
\item Delayed halo collapse, implies that at $z=0$ WDM haloes are smaller than their CDM counterparts.
\item Baryons are more centrally concentrated in CDM versus WDM haloes.
\item In one of the galaxies we simulated, a clearly identifiable disc is found. This is fatter and smaller in WDM, a consequence of it being younger and having more recent merger activity.
\end{itemize}
 
Our conclusions are all consequences of the delayed formation and collapse of haloes in WDM cosmologies with respect to CDM. This simple attribute, a direct result of the lack of small scale power due to free streaming of DM at early times, informs a myriad of physical properties, from star formation rates to bulge/disc ratios to colors. One of the more intriguing findings in this work is the thickening of the one disc we formed (in halo C) in our WDM run. It remains to be seen if this is simply due to the unique dynamical history of this particular realization or if WDM generically and systematically produces thicker discs than CDM.

\section*{Acknowledgments} 
NIL is supported through a grant from the Deustche Forschungs Gemeinschaft (DFG). ADC acknowledges the AIP - Leibniz-Institut f\"{u}r Astrophysik, where this work has been partially carried on. AK is supported by the {\it Spanish Ministerio de Ciencia e Innovaci\'on} (MICINN) in Spain through the Ramon y Cajal programme as well as the grants AYA 2009-13875-C03-02, AYA2009-12792-C03-03, CSD2009-00064, and CAM S2009/ESP-1496 and the {\it Ministerio de Econom\'ia y Competitividad} (MINECO) through grant AYA2012-31101. GY also acknowledges support from MINECO through research projects 
 CSD2007-0050, AYA 2009-13875-C03-02 and AYA 2012-31101, and from Comunidad de Madrid  through  ASTROMADRID project (CAM S2009/ESP-1496). YH has been partially supported by ISF 1013/12. The simulations were performed and analyzed at  the Leibniz Rechenzentrum Munich (LRZ) and at the Barcelona Supercomputing Center (BSC).

\bibliography{./ref}

\begin{thebibliography}{60}
\expandafter\ifx\csname natexlab\endcsname\relax\def\natexlab#1{#1}\fi
\expandafter\ifx\csname url\endcsname\relax
  \def\url#1{\texttt{#1}}\fi
\expandafter\ifx\csname urlprefix\endcsname\relax\def\urlprefix{URL }\fi
\providecommand{\eprint}[2][]{\url{#2}}
\providecommand{\bibinfo}[2]{#2}
\ifx\xfnm\relax \def\xfnm[#1]{\unskip,\space#1}\fi
%Type = Article
\bibitem[{{Abadi} et~al.(2003){Abadi}, {Navarro}, {Steinmetz} and
  {Eke}}]{Abadi03b}
\bibinfo{author}{{Abadi}, M.G.}, \bibinfo{author}{{Navarro}, J.F.},
  \bibinfo{author}{{Steinmetz}, M.}, \bibinfo{author}{{Eke}, V.R.},
  \bibinfo{year}{2003}.
\newblock \bibinfo{title}{{Simulations of Galaxy Formation in a {$\Lambda$}
  Cold Dark Matter Universe. II. The Fine Structure of Simulated Galactic
  Disks}}.
\newblock \bibinfo{journal}{\apj} \bibinfo{volume}{597},
  \bibinfo{pages}{21--34}.
\newblock \eprint{arXiv:astro-ph/0212282}.
%Type = Article
\bibitem[{{Angulo} et~al.(2012){Angulo}, {Springel}, {White}, {Jenkins},
  {Baugh} and {Frenk}}]{2012MNRAS.426.2046A}
\bibinfo{author}{{Angulo}, R.E.}, \bibinfo{author}{{Springel}, V.},
  \bibinfo{author}{{White}, S.D.M.}, \bibinfo{author}{{Jenkins}, A.},
  \bibinfo{author}{{Baugh}, C.M.}, \bibinfo{author}{{Frenk}, C.S.},
  \bibinfo{year}{2012}.
\newblock \bibinfo{title}{{Scaling relations for galaxy clusters in the
  Millennium-XXL simulation}}.
\newblock \bibinfo{journal}{\mnras} \bibinfo{volume}{426},
  \bibinfo{pages}{2046--2062}.
\newblock \eprint{1203.3216}.
%Type = Article
\bibitem[{{Avila-Reese} et~al.(2001){Avila-Reese}, {Col{\'{\i}}n},
  {Valenzuela}, {D'Onghia} and {Firmani}}]{2001ApJ...559..516A}
\bibinfo{author}{{Avila-Reese}, V.}, \bibinfo{author}{{Col{\'{\i}}n}, P.},
  \bibinfo{author}{{Valenzuela}, O.}, \bibinfo{author}{{D'Onghia}, E.},
  \bibinfo{author}{{Firmani}, C.}, \bibinfo{year}{2001}.
\newblock \bibinfo{title}{{Formation and Structure of Halos in a Warm Dark
  Matter Cosmology}}.
\newblock \bibinfo{journal}{\apj} \bibinfo{volume}{559},
  \bibinfo{pages}{516--530}.
\newblock \eprint{arXiv:astro-ph/0010525}.
%Type = Article
\bibitem[{{Benson} et~al.(2003){Benson}, {Bower}, {Frenk}, {Lacey}, {Baugh} and
  {Cole}}]{2003ApJ...599...38B}
\bibinfo{author}{{Benson}, A.J.}, \bibinfo{author}{{Bower}, R.G.},
  \bibinfo{author}{{Frenk}, C.S.}, \bibinfo{author}{{Lacey}, C.G.},
  \bibinfo{author}{{Baugh}, C.M.}, \bibinfo{author}{{Cole}, S.},
  \bibinfo{year}{2003}.
\newblock \bibinfo{title}{{What Shapes the Luminosity Function of Galaxies?}}
\newblock \bibinfo{journal}{\apj} \bibinfo{volume}{599},
  \bibinfo{pages}{38--49}.
\newblock \eprint{arXiv:astro-ph/0302450}.
%Type = Article
\bibitem[{{Bertone} et~al.(2005){Bertone}, {Hooper} and
  {Silk}}]{2005PhR...405..279B}
\bibinfo{author}{{Bertone}, G.}, \bibinfo{author}{{Hooper}, D.},
  \bibinfo{author}{{Silk}, J.}, \bibinfo{year}{2005}.
\newblock \bibinfo{title}{{Particle dark matter: evidence, candidates and
  constraints}}.
\newblock \bibinfo{journal}{\physrep} \bibinfo{volume}{405},
  \bibinfo{pages}{279--390}.
\newblock \eprint{arXiv:hep-ph/0404175}.
%Type = Article
\bibitem[{{Bode} et~al.(2001){Bode}, {Ostriker} and {Turok}}]{bode2001}
\bibinfo{author}{{Bode}, P.}, \bibinfo{author}{{Ostriker}, J.P.},
  \bibinfo{author}{{Turok}, N.}, \bibinfo{year}{2001}.
\newblock \bibinfo{title}{{Halo Formation in Warm Dark Matter Models}}.
\newblock \bibinfo{journal}{\apj} \bibinfo{volume}{556},
  \bibinfo{pages}{93--107}.
\newblock \eprint{arXiv:astro-ph/0010389}.
%Type = Article
\bibitem[{{Boyarsky} et~al.(2009a){Boyarsky}, {Lesgourgues}, {Ruchayskiy} and
  {Viel}}]{2009PhRvL.102t1304B}
\bibinfo{author}{{Boyarsky}, A.}, \bibinfo{author}{{Lesgourgues}, J.},
  \bibinfo{author}{{Ruchayskiy}, O.}, \bibinfo{author}{{Viel}, M.},
  \bibinfo{year}{2009}a.
\newblock \bibinfo{title}{{Realistic Sterile Neutrino Dark Matter with KeV Mass
  does not Contradict Cosmological Bounds}}.
\newblock \bibinfo{journal}{Physical Review Letters} \bibinfo{volume}{102},
  \bibinfo{pages}{201304}.
\newblock \eprint{0812.3256}.
%Type = Article
\bibitem[{{Boyarsky} et~al.(2009b){Boyarsky}, {Ruchayskiy} and
  {Shaposhnikov}}]{2009ARNPS..59..191B}
\bibinfo{author}{{Boyarsky}, A.}, \bibinfo{author}{{Ruchayskiy}, O.},
  \bibinfo{author}{{Shaposhnikov}, M.}, \bibinfo{year}{2009}b.
\newblock \bibinfo{title}{{The Role of Sterile Neutrinos in Cosmology and
  Astrophysics}}.
\newblock \bibinfo{journal}{Annual Review of Nuclear and Particle Science}
  \bibinfo{volume}{59}, \bibinfo{pages}{191--214}.
\newblock \eprint{0901.0011}.
%Type = Article
\bibitem[{{Boylan-Kolchin} et~al.(2011){Boylan-Kolchin}, {Bullock} and
  {Kaplinghat}}]{Boylan11}
\bibinfo{author}{{Boylan-Kolchin}, M.}, \bibinfo{author}{{Bullock}, J.S.},
  \bibinfo{author}{{Kaplinghat}, M.}, \bibinfo{year}{2011}.
\newblock \bibinfo{title}{{Too big to fail? The puzzling darkness of massive
  Milky Way subhaloes}}.
\newblock \bibinfo{journal}{\mnras} \bibinfo{volume}{415},
  \bibinfo{pages}{L40--L44}.
\newblock \eprint{1103.0007}.
%Type = Article
\bibitem[{{Boylan-Kolchin} et~al.(2012){Boylan-Kolchin}, {Bullock} and
  {Kaplinghat}}]{Boylan12}
\bibinfo{author}{{Boylan-Kolchin}, M.}, \bibinfo{author}{{Bullock}, J.S.},
  \bibinfo{author}{{Kaplinghat}, M.}, \bibinfo{year}{2012}.
\newblock \bibinfo{title}{{The Milky Way's bright satellites as an apparent
  failure of {$\Lambda$}CDM}}.
\newblock \bibinfo{journal}{\mnras} , \bibinfo{pages}{2657}\eprint{1111.2048}.
%Type = Article
\bibitem[{{Busha} et~al.(2011){Busha}, {Wechsler}, {Behroozi}, {Gerke},
  {Klypin} and {Primack}}]{2011ApJ...743..117B}
\bibinfo{author}{{Busha}, M.T.}, \bibinfo{author}{{Wechsler}, R.H.},
  \bibinfo{author}{{Behroozi}, P.S.}, \bibinfo{author}{{Gerke}, B.F.},
  \bibinfo{author}{{Klypin}, A.A.}, \bibinfo{author}{{Primack}, J.R.},
  \bibinfo{year}{2011}.
\newblock \bibinfo{title}{{Statistics of Satellite Galaxies around
  Milky-Way-like Hosts}}.
\newblock \bibinfo{journal}{\apj} \bibinfo{volume}{743}, \bibinfo{pages}{117}.
\newblock \eprint{1011.6373}.
%Type = Article
\bibitem[{{Dayal} and {Libeskind}(2012)}]{2012MNRAS.419L...9D}
\bibinfo{author}{{Dayal}, P.}, \bibinfo{author}{{Libeskind}, N.I.},
  \bibinfo{year}{2012}.
\newblock \bibinfo{title}{{Local Group progenitors: Lyman Alpha bright?}}
\newblock \bibinfo{journal}{\mnras} \bibinfo{volume}{419},
  \bibinfo{pages}{L9--L13}.
\newblock \eprint{1107.5721}.
%Type = Article
\bibitem[{{de Rossi} et~al.(2009){de Rossi}, {Tissera}, {De Lucia} and
  {Kauffmann}}]{2009MNRAS.395..210D}
\bibinfo{author}{{de Rossi}, M.E.}, \bibinfo{author}{{Tissera}, P.B.},
  \bibinfo{author}{{De Lucia}, G.}, \bibinfo{author}{{Kauffmann}, G.},
  \bibinfo{year}{2009}.
\newblock \bibinfo{title}{{Milky Way type galaxies in a {$\Lambda$}CDM
  cosmology}}.
\newblock \bibinfo{journal}{\mnras} \bibinfo{volume}{395},
  \bibinfo{pages}{210--217}.
\newblock \eprint{0806.2872}.
%Type = Article
\bibitem[{{Di Cintio} et~al.(2012a){Di Cintio}, {Knebe}, {Libeskind}, {Brook},
  {Yepes}, {Gottloeber} and {Hoffman}}]{DiCintio12}
\bibinfo{author}{{Di Cintio}, A.}, \bibinfo{author}{{Knebe}, A.},
  \bibinfo{author}{{Libeskind}, N.I.}, \bibinfo{author}{{Brook}, C.},
  \bibinfo{author}{{Yepes}, G.}, \bibinfo{author}{{Gottloeber}, S.},
  \bibinfo{author}{{Hoffman}, Y.}, \bibinfo{year}{2012}a.
\newblock \bibinfo{title}{{Size matters: the non-universal density profile of
  subhaloes in SPH simulations and implications for the Milky Way's dSphs}}.
\newblock \bibinfo{journal}{ArXiv e-prints} \eprint{1204.0515}.
%Type = Article
\bibitem[{{Di Cintio} et~al.(2012b){Di Cintio}, {Knebe}, {Libeskind},
  {Hoffman}, {Yepes} and {Gottl{\"o}ber}}]{DiCintio12b}
\bibinfo{author}{{Di Cintio}, A.}, \bibinfo{author}{{Knebe}, A.},
  \bibinfo{author}{{Libeskind}, N.I.}, \bibinfo{author}{{Hoffman}, Y.},
  \bibinfo{author}{{Yepes}, G.}, \bibinfo{author}{{Gottl{\"o}ber}, S.},
  \bibinfo{year}{2012}b.
\newblock \bibinfo{title}{{Applying scale-free mass estimators to the Local
  Group in Constrained Local Universe Simulations}}.
\newblock \bibinfo{journal}{\mnras} \bibinfo{volume}{423},
  \bibinfo{pages}{1883--1895}.
\newblock \eprint{1204.0005}.
%Type = Article
\bibitem[{{Di Cintio} et~al.(2011){Di Cintio}, {Knebe}, {Libeskind}, {Yepes},
  {Gottl{\"o}ber} and {Hoffman}}]{DiCintio11}
\bibinfo{author}{{Di Cintio}, A.}, \bibinfo{author}{{Knebe}, A.},
  \bibinfo{author}{{Libeskind}, N.I.}, \bibinfo{author}{{Yepes}, G.},
  \bibinfo{author}{{Gottl{\"o}ber}, S.}, \bibinfo{author}{{Hoffman}, Y.},
  \bibinfo{year}{2011}.
\newblock \bibinfo{title}{{Too small to succeed? Lighting up massive dark
  matter subhaloes of the Milky Way}}.
\newblock \bibinfo{journal}{\mnras} \bibinfo{volume}{417},
  \bibinfo{pages}{L74--L78}.
\newblock \eprint{1107.5045}.
%Type = Article
\bibitem[{{Forero-Romero} et~al.(2011){Forero-Romero}, {Hoffman}, {Yepes},
  {Gottoeber}, {Piontek}, {Klypin} and {Steinmetz}}]{2011arXiv1107.0017F}
\bibinfo{author}{{Forero-Romero}, J.E.}, \bibinfo{author}{{Hoffman}, Y.},
  \bibinfo{author}{{Yepes}, G.}, \bibinfo{author}{{Gottoeber}, S.},
  \bibinfo{author}{{Piontek}, R.}, \bibinfo{author}{{Klypin}, A.},
  \bibinfo{author}{{Steinmetz}, M.}, \bibinfo{year}{2011}.
\newblock \bibinfo{title}{{The dark matter assembly of the Local Group in
  constrained cosmological simulations of a LambdaCDM universe}}.
\newblock \bibinfo{journal}{ArXiv e-prints} \eprint{1107.0017}.
%Type = Article
\bibitem[{{Gao} et~al.(2012){Gao}, {Navarro}, {Frenk}, {Jenkins}, {Springel}
  and {White}}]{2012MNRAS.425.2169G}
\bibinfo{author}{{Gao}, L.}, \bibinfo{author}{{Navarro}, J.F.},
  \bibinfo{author}{{Frenk}, C.S.}, \bibinfo{author}{{Jenkins}, A.},
  \bibinfo{author}{{Springel}, V.}, \bibinfo{author}{{White}, S.D.M.},
  \bibinfo{year}{2012}.
\newblock \bibinfo{title}{{The Phoenix Project: the dark side of rich Galaxy
  clusters}}.
\newblock \bibinfo{journal}{\mnras} \bibinfo{volume}{425},
  \bibinfo{pages}{2169--2186}.
\newblock \eprint{1201.1940}.
%Type = Article
\bibitem[{{Gill} et~al.(2004){Gill}, {Knebe}, {Gibson} and {Dopita}}]{GIll04b}
\bibinfo{author}{{Gill}, S.P.D.}, \bibinfo{author}{{Knebe}, A.},
  \bibinfo{author}{{Gibson}, B.K.}, \bibinfo{author}{{Dopita}, M.A.},
  \bibinfo{year}{2004}.
\newblock \bibinfo{title}{{The evolution of substructure - II. Linking dynamics
  to environment}}.
\newblock \bibinfo{journal}{\mnras} \bibinfo{volume}{351},
  \bibinfo{pages}{410--422}.
\newblock \eprint{arXiv:astro-ph/0404255}.
%Type = Article
\bibitem[{{Haardt} and {Madau}(1996)}]{1996ApJ...461...20H}
\bibinfo{author}{{Haardt}, F.}, \bibinfo{author}{{Madau}, P.},
  \bibinfo{year}{1996}.
\newblock \bibinfo{title}{{Radiative Transfer in a Clumpy Universe. II. The
  Ultraviolet Extragalactic Background}}.
\newblock \bibinfo{journal}{\apj} \bibinfo{volume}{461},
  \bibinfo{pages}{20--+}.
\newblock \eprint{arXiv:astro-ph/9509093}.
%Type = Article
\bibitem[{{Hoffman} and {Ribak}(1991)}]{1991ApJ...380L...5H}
\bibinfo{author}{{Hoffman}, Y.}, \bibinfo{author}{{Ribak}, E.},
  \bibinfo{year}{1991}.
\newblock \bibinfo{title}{{Constrained realizations of Gaussian fields - A
  simple algorithm}}.
\newblock \bibinfo{journal}{\apjl} \bibinfo{volume}{380},
  \bibinfo{pages}{L5--L8}.
%Type = Article
\bibitem[{{Ishiyama} et~al.(2010){Ishiyama}, {Makino} and
  {Ebisuzaki}}]{2010ApJ...723L.195I}
\bibinfo{author}{{Ishiyama}, T.}, \bibinfo{author}{{Makino}, J.},
  \bibinfo{author}{{Ebisuzaki}, T.}, \bibinfo{year}{2010}.
\newblock \bibinfo{title}{{Gamma-ray Signal from Earth-mass Dark Matter
  Microhalos}}.
\newblock \bibinfo{journal}{\apjl} \bibinfo{volume}{723},
  \bibinfo{pages}{L195--L200}.
\newblock \eprint{1006.3392}.
%Type = Article
\bibitem[{{Karachentsev} et~al.(2004){Karachentsev}, {Karachentseva},
  {Huchtmeier} and {Makarov}}]{2004AJ....127.2031K}
\bibinfo{author}{{Karachentsev}, I.D.}, \bibinfo{author}{{Karachentseva},
  V.E.}, \bibinfo{author}{{Huchtmeier}, W.K.}, \bibinfo{author}{{Makarov},
  D.I.}, \bibinfo{year}{2004}.
\newblock \bibinfo{title}{{A Catalog of Neighboring Galaxies}}.
\newblock \bibinfo{journal}{\aj} \bibinfo{volume}{127},
  \bibinfo{pages}{2031--2068}.
%Type = Article
\bibitem[{{Kim} et~al.(2011){Kim}, {Park}, {Rossi}, {Lee} and
  {Gott}}]{2011JKAS...44..217K}
\bibinfo{author}{{Kim}, J.}, \bibinfo{author}{{Park}, C.},
  \bibinfo{author}{{Rossi}, G.}, \bibinfo{author}{{Lee}, S.M.},
  \bibinfo{author}{{Gott}, III, J.R.}, \bibinfo{year}{2011}.
\newblock \bibinfo{title}{{The New Horizon Run Cosmological N-Body
  Simulations}}.
\newblock \bibinfo{journal}{Journal of Korean Astronomical Society}
  \bibinfo{volume}{44}, \bibinfo{pages}{217--234}.
\newblock \eprint{1112.1754}.
%Type = Article
\bibitem[{{Klypin} et~al.(2001){Klypin}, {Kravtsov}, {Bullock} and
  {Primack}}]{2001ApJ...554..903K}
\bibinfo{author}{{Klypin}, A.}, \bibinfo{author}{{Kravtsov}, A.V.},
  \bibinfo{author}{{Bullock}, J.S.}, \bibinfo{author}{{Primack}, J.R.},
  \bibinfo{year}{2001}.
\newblock \bibinfo{title}{{Resolving the Structure of Cold Dark Matter Halos}}.
\newblock \bibinfo{journal}{\apj} \bibinfo{volume}{554},
  \bibinfo{pages}{903--915}.
\newblock \eprint{arXiv:astro-ph/0006343}.
%Type = Article
\bibitem[{{Klypin} et~al.(1999){Klypin}, {Kravtsov}, {Valenzuela} and
  {Prada}}]{1999ApJ...522...82K}
\bibinfo{author}{{Klypin}, A.}, \bibinfo{author}{{Kravtsov}, A.V.},
  \bibinfo{author}{{Valenzuela}, O.}, \bibinfo{author}{{Prada}, F.},
  \bibinfo{year}{1999}.
\newblock \bibinfo{title}{{Where Are the Missing Galactic Satellites?}}
\newblock \bibinfo{journal}{\apj} \bibinfo{volume}{522},
  \bibinfo{pages}{82--92}.
\newblock \eprint{arXiv:astro-ph/9901240}.
%Type = Article
\bibitem[{{Knebe} et~al.(2008){Knebe}, {Arnold}, {Power} and
  {Gibson}}]{2008MNRAS.386.1029K}
\bibinfo{author}{{Knebe}, A.}, \bibinfo{author}{{Arnold}, B.},
  \bibinfo{author}{{Power}, C.}, \bibinfo{author}{{Gibson}, B.K.},
  \bibinfo{year}{2008}.
\newblock \bibinfo{title}{{The dynamics of subhaloes in warm dark matter
  models}}.
\newblock \bibinfo{journal}{\mnras} \bibinfo{volume}{386},
  \bibinfo{pages}{1029--1037}.
\newblock \eprint{0802.1628}.
%Type = Article
\bibitem[{{Knebe} et~al.(2002){Knebe}, {Devriendt}, {Mahmood} and
  {Silk}}]{2002MNRAS.329..813K}
\bibinfo{author}{{Knebe}, A.}, \bibinfo{author}{{Devriendt}, J.E.G.},
  \bibinfo{author}{{Mahmood}, A.}, \bibinfo{author}{{Silk}, J.},
  \bibinfo{year}{2002}.
\newblock \bibinfo{title}{{Merger histories in warm dark matter structure
  formation scenarios}}.
\newblock \bibinfo{journal}{\mnras} \bibinfo{volume}{329},
  \bibinfo{pages}{813--828}.
\newblock \eprint{arXiv:astro-ph/0105316}.
%Type = Article
\bibitem[{{Knebe} et~al.(2011a){Knebe}, {Libeskind}, {Doumler}, {Yepes},
  {Gottloeber} and {Hoffman}}]{2011arXiv1107.2944K}
\bibinfo{author}{{Knebe}, A.}, \bibinfo{author}{{Libeskind}, N.I.},
  \bibinfo{author}{{Doumler}, T.}, \bibinfo{author}{{Yepes}, G.},
  \bibinfo{author}{{Gottloeber}, S.}, \bibinfo{author}{{Hoffman}, Y.},
  \bibinfo{year}{2011}a.
\newblock \bibinfo{title}{{Renegade Subhaloes in the Local Group}}.
\newblock \bibinfo{journal}{ArXiv e-prints} \eprint{1107.2944}.
%Type = Article
\bibitem[{{Knebe} et~al.(2011b){Knebe}, {Libeskind}, {Knollmann},
  {Martinez-Vaquero}, {Yepes}, {Gottl{\"o}ber} and
  {Hoffman}}]{2011MNRAS.412..529K}
\bibinfo{author}{{Knebe}, A.}, \bibinfo{author}{{Libeskind}, N.I.},
  \bibinfo{author}{{Knollmann}, S.R.}, \bibinfo{author}{{Martinez-Vaquero},
  L.A.}, \bibinfo{author}{{Yepes}, G.}, \bibinfo{author}{{Gottl{\"o}ber}, S.},
  \bibinfo{author}{{Hoffman}, Y.}, \bibinfo{year}{2011}b.
\newblock \bibinfo{title}{{The luminosities of backsplash galaxies in
  constrained simulations of the Local Group}}.
\newblock \bibinfo{journal}{\mnras} \bibinfo{volume}{412},
  \bibinfo{pages}{529--536}.
\newblock \eprint{1010.5670}.
%Type = Article
\bibitem[{{Knebe} et~al.(2010){Knebe}, {Libeskind}, {Knollmann}, {Yepes},
  {Gottl{\"o}ber} and {Hoffman}}]{2010MNRAS.405.1119K}
\bibinfo{author}{{Knebe}, A.}, \bibinfo{author}{{Libeskind}, N.I.},
  \bibinfo{author}{{Knollmann}, S.R.}, \bibinfo{author}{{Yepes}, G.},
  \bibinfo{author}{{Gottl{\"o}ber}, S.}, \bibinfo{author}{{Hoffman}, Y.},
  \bibinfo{year}{2010}.
\newblock \bibinfo{title}{{The impact of baryonic physics on the shape and
  radial alignment of substructures in cosmological dark matter haloes}}.
\newblock \bibinfo{journal}{\mnras} \bibinfo{volume}{405},
  \bibinfo{pages}{1119--1128}.
\newblock \eprint{1002.2853}.
%Type = Article
\bibitem[{{Knebe} et~al.(2013){Knebe}, {Libeskind}, {Pearce}, {Behroozi},
  {Casado}, {Dolag}, {Dominguez-Tenreiro}, {Elahi}, {Lux}, {Muldrew} and
  {Onions}}]{2013MNRAS.428.2039K}
\bibinfo{author}{{Knebe}, A.}, \bibinfo{author}{{Libeskind}, N.I.},
  \bibinfo{author}{{Pearce}, F.}, \bibinfo{author}{{Behroozi}, P.},
  \bibinfo{author}{{Casado}, J.}, \bibinfo{author}{{Dolag}, K.},
  \bibinfo{author}{{Dominguez-Tenreiro}, R.}, \bibinfo{author}{{Elahi}, P.},
  \bibinfo{author}{{Lux}, H.}, \bibinfo{author}{{Muldrew}, S.I.},
  \bibinfo{author}{{Onions}, J.}, \bibinfo{year}{2013}.
\newblock \bibinfo{title}{{Galaxies going MAD: the Galaxy-Finder Comparison
  Project}}.
\newblock \bibinfo{journal}{\mnras} \bibinfo{volume}{428},
  \bibinfo{pages}{2039--2052}.
\newblock \eprint{1210.2578}.
%Type = Article
\bibitem[{{Knollmann} and {Knebe}(2009)}]{2009ApJS..182..608K}
\bibinfo{author}{{Knollmann}, S.R.}, \bibinfo{author}{{Knebe}, A.},
  \bibinfo{year}{2009}.
\newblock \bibinfo{title}{{AHF: Amiga's Halo Finder}}.
\newblock \bibinfo{journal}{\apjs} \bibinfo{volume}{182},
  \bibinfo{pages}{608--624}.
\newblock \eprint{0904.3662}.
%Type = Article
\bibitem[{{Libeskind} et~al.(2005){Libeskind}, {Frenk}, {Cole}, {Helly},
  {Jenkins}, {Navarro} and {Power}}]{2005MNRAS.363..146L}
\bibinfo{author}{{Libeskind}, N.I.}, \bibinfo{author}{{Frenk}, C.S.},
  \bibinfo{author}{{Cole}, S.}, \bibinfo{author}{{Helly}, J.C.},
  \bibinfo{author}{{Jenkins}, A.}, \bibinfo{author}{{Navarro}, J.F.},
  \bibinfo{author}{{Power}, C.}, \bibinfo{year}{2005}.
\newblock \bibinfo{title}{{The distribution of satellite galaxies: the great
  pancake}}.
\newblock \bibinfo{journal}{\mnras} \bibinfo{volume}{363},
  \bibinfo{pages}{146--152}.
\newblock \eprint{arXiv:astro-ph/0503400}.
%Type = Article
\bibitem[{{Libeskind} et~al.(2011a){Libeskind}, {Knebe}, {Hoffman},
  {Gottl{\"o}ber}, {Yepes} and {Steinmetz}}]{2011MNRAS.411.1525L}
\bibinfo{author}{{Libeskind}, N.I.}, \bibinfo{author}{{Knebe}, A.},
  \bibinfo{author}{{Hoffman}, Y.}, \bibinfo{author}{{Gottl{\"o}ber}, S.},
  \bibinfo{author}{{Yepes}, G.}, \bibinfo{author}{{Steinmetz}, M.},
  \bibinfo{year}{2011}a.
\newblock \bibinfo{title}{{The preferred direction of infalling satellite
  galaxies in the Local Group}}.
\newblock \bibinfo{journal}{\mnras} \bibinfo{volume}{411},
  \bibinfo{pages}{1525--1535}.
\newblock \eprint{1010.1531}.
%Type = Article
\bibitem[{{Libeskind} et~al.(2011b){Libeskind}, {Knebe}, {Hoffman} and
  {Gottloeber}}]{Libeskindetal2011}
\bibinfo{author}{{Libeskind}, N.I.}, \bibinfo{author}{{Knebe}, A.},
  \bibinfo{author}{{Hoffman}, Y.}, \bibinfo{author}{{Gottloeber}, S.~{Yepes},
  G.a.}, \bibinfo{year}{2011}b.
\newblock \bibinfo{title}{{Disentangling the dark matter from the stellar
  halo}}.
\newblock \bibinfo{journal}{ArXiv e-prints} \eprint{n/a}.
%Type = Article
\bibitem[{{Libeskind} et~al.(2010){Libeskind}, {Yepes}, {Knebe},
  {Gottl{\"o}ber}, {Hoffman} and {Knollmann}}]{2010MNRAS.401.1889L}
\bibinfo{author}{{Libeskind}, N.I.}, \bibinfo{author}{{Yepes}, G.},
  \bibinfo{author}{{Knebe}, A.}, \bibinfo{author}{{Gottl{\"o}ber}, S.},
  \bibinfo{author}{{Hoffman}, Y.}, \bibinfo{author}{{Knollmann}, S.R.},
  \bibinfo{year}{2010}.
\newblock \bibinfo{title}{{Constrained simulations of the Local Group: on the
  radial distribution of substructures}}.
\newblock \bibinfo{journal}{\mnras} \bibinfo{volume}{401},
  \bibinfo{pages}{1889--1897}.
\newblock \eprint{0909.4423}.
%Type = Article
\bibitem[{{Lovell} et~al.(2011){Lovell}, {Eke}, {Frenk}, {Gao}, {Jenkins},
  {Theuns}, {Wang}, {Boyarsky} and {Ruchayskiy}}]{Lovell11}
\bibinfo{author}{{Lovell}, M.}, \bibinfo{author}{{Eke}, V.},
  \bibinfo{author}{{Frenk}, C.}, \bibinfo{author}{{Gao}, L.},
  \bibinfo{author}{{Jenkins}, A.}, \bibinfo{author}{{Theuns}, T.},
  \bibinfo{author}{{Wang}, J.}, \bibinfo{author}{{Boyarsky}, A.},
  \bibinfo{author}{{Ruchayskiy}, O.}, \bibinfo{year}{2011}.
\newblock \bibinfo{title}{{The Haloes of Bright Satellite Galaxies in a Warm
  Dark Matter Universe}}.
\newblock \bibinfo{journal}{ArXiv e-prints} \eprint{1104.2929}.
%Type = Article
\bibitem[{{Macci{\`o}} and {Fontanot}(2010)}]{2010MNRAS.404L..16M}
\bibinfo{author}{{Macci{\`o}}, A.V.}, \bibinfo{author}{{Fontanot}, F.},
  \bibinfo{year}{2010}.
\newblock \bibinfo{title}{{How cold is dark matter? Constraints from Milky Way
  satellites}}.
\newblock \bibinfo{journal}{\mnras} \bibinfo{volume}{404},
  \bibinfo{pages}{L16--L20}.
\newblock \eprint{0910.2460}.
%Type = Article
\bibitem[{{Macci{\`o}} et~al.(2013){Macci{\`o}}, {Ruchayskiy}, {Boyarsky} and
  {Mu{\~n}oz-Cuartas}}]{2013MNRAS.428..882M}
\bibinfo{author}{{Macci{\`o}}, A.V.}, \bibinfo{author}{{Ruchayskiy}, O.},
  \bibinfo{author}{{Boyarsky}, A.}, \bibinfo{author}{{Mu{\~n}oz-Cuartas},
  J.C.}, \bibinfo{year}{2013}.
\newblock \bibinfo{title}{{The inner structure of haloes in cold+warm dark
  matter models}}.
\newblock \bibinfo{journal}{\mnras} \bibinfo{volume}{428},
  \bibinfo{pages}{882--890}.
\newblock \eprint{1202.2858}.
%Type = Article
\bibitem[{{Moore} et~al.(1999){Moore}, {Ghigna}, {Governato}, {Lake}, {Quinn},
  {Stadel} and {Tozzi}}]{1999ApJ...524L..19M}
\bibinfo{author}{{Moore}, B.}, \bibinfo{author}{{Ghigna}, S.},
  \bibinfo{author}{{Governato}, F.}, \bibinfo{author}{{Lake}, G.},
  \bibinfo{author}{{Quinn}, T.}, \bibinfo{author}{{Stadel}, J.},
  \bibinfo{author}{{Tozzi}, P.}, \bibinfo{year}{1999}.
\newblock \bibinfo{title}{{Dark Matter Substructure within Galactic Halos}}.
\newblock \bibinfo{journal}{\apjl} \bibinfo{volume}{524},
  \bibinfo{pages}{L19--L22}.
\newblock \eprint{arXiv:astro-ph/9907411}.
%Type = Article
\bibitem[{{Reiprich} and {B{\"o}hringer}(2002)}]{2002ApJ...567..716R}
\bibinfo{author}{{Reiprich}, T.H.}, \bibinfo{author}{{B{\"o}hringer}, H.},
  \bibinfo{year}{2002}.
\newblock \bibinfo{title}{{The Mass Function of an X-Ray Flux-limited Sample of
  Galaxy Clusters}}.
\newblock \bibinfo{journal}{\apj} \bibinfo{volume}{567},
  \bibinfo{pages}{716--740}.
\newblock \eprint{arXiv:astro-ph/0111285}.
%Type = Article
\bibitem[{{Riebe} et~al.(2011){Riebe}, {Partl}, {Enke}, {Forero-Romero},
  {Gottloeber}, {Klypin}, {Lemson}, {Prada}, {Primack}, {Steinmetz} and
  {Turchaninov}}]{2011arXiv1109.0003R}
\bibinfo{author}{{Riebe}, K.}, \bibinfo{author}{{Partl}, A.M.},
  \bibinfo{author}{{Enke}, H.}, \bibinfo{author}{{Forero-Romero}, J.},
  \bibinfo{author}{{Gottloeber}, S.}, \bibinfo{author}{{Klypin}, A.},
  \bibinfo{author}{{Lemson}, G.}, \bibinfo{author}{{Prada}, F.},
  \bibinfo{author}{{Primack}, J.R.}, \bibinfo{author}{{Steinmetz}, M.},
  \bibinfo{author}{{Turchaninov}, V.}, \bibinfo{year}{2011}.
\newblock \bibinfo{title}{{The MultiDark Database: Release of the Bolshoi and
  MultiDark Cosmological Simulations}}.
\newblock \bibinfo{journal}{ArXiv e-prints} \eprint{1109.0003}.
%Type = Article
\bibitem[{{Sales} et~al.(2012){Sales}, {Navarro}, {Theuns}, {Schaye}, {White},
  {Frenk}, {Crain} and {Dalla Vecchia}}]{Sales12}
\bibinfo{author}{{Sales}, L.V.}, \bibinfo{author}{{Navarro}, J.F.},
  \bibinfo{author}{{Theuns}, T.}, \bibinfo{author}{{Schaye}, J.},
  \bibinfo{author}{{White}, S.D.M.}, \bibinfo{author}{{Frenk}, C.S.},
  \bibinfo{author}{{Crain}, R.A.}, \bibinfo{author}{{Dalla Vecchia}, C.},
  \bibinfo{year}{2012}.
\newblock \bibinfo{title}{{The origin of discs and spheroids in simulated
  galaxies}}.
\newblock \bibinfo{journal}{\mnras} \bibinfo{volume}{423},
  \bibinfo{pages}{1544--1555}.
\newblock \eprint{1112.2220}.
%Type = Article
\bibitem[{{Scannapieco} et~al.(2010){Scannapieco}, {Gadotti}, {Jonsson} and
  {White}}]{Scannapieco10}
\bibinfo{author}{{Scannapieco}, C.}, \bibinfo{author}{{Gadotti}, D.A.},
  \bibinfo{author}{{Jonsson}, P.}, \bibinfo{author}{{White}, S.D.M.},
  \bibinfo{year}{2010}.
\newblock \bibinfo{title}{{An observer's view of simulated galaxies:
  disc-to-total ratios, bars and (pseudo-)bulges}}.
\newblock \bibinfo{journal}{\mnras} \bibinfo{volume}{407},
  \bibinfo{pages}{L41--L45}.
\newblock \eprint{1001.4890}.
%Type = Article
\bibitem[{{Spergel} et~al.(2007){Spergel}, {Bean}, {Dor{\'e}}, {Nolta},
  {Bennett}, {Dunkley}, {Hinshaw}, {Jarosik} and {et
  al.}}]{2007ApJS..170..377S}
\bibinfo{author}{{Spergel}, D.N.}, \bibinfo{author}{{Bean}, R.},
  \bibinfo{author}{{Dor{\'e}}, O.}, \bibinfo{author}{{Nolta}, M.R.},
  \bibinfo{author}{{Bennett}, C.L.}, \bibinfo{author}{{Dunkley}, J.},
  \bibinfo{author}{{Hinshaw}, G.}, \bibinfo{author}{{Jarosik}, N.},
  \bibinfo{author}{{et al.}}, \bibinfo{year}{2007}.
\newblock \bibinfo{title}{{Three-Year Wilkinson Microwave Anisotropy Probe
  (WMAP) Observations: Implications for Cosmology}}.
\newblock \bibinfo{journal}{\apjs} \bibinfo{volume}{170},
  \bibinfo{pages}{377--408}.
\newblock \eprint{arXiv:astro-ph/0603449}.
%Type = Article
\bibitem[{{Springel}(2005)}]{2005MNRAS.364.1105S}
\bibinfo{author}{{Springel}, V.}, \bibinfo{year}{2005}.
\newblock \bibinfo{title}{{The cosmological simulation code GADGET-2}}.
\newblock \bibinfo{journal}{\mnras} \bibinfo{volume}{364},
  \bibinfo{pages}{1105--1134}.
\newblock \eprint{arXiv:astro-ph/0505010}.
%Type = Article
\bibitem[{{Springel} and {Hernquist}(2003)}]{2003MNRAS.339..289S}
\bibinfo{author}{{Springel}, V.}, \bibinfo{author}{{Hernquist}, L.},
  \bibinfo{year}{2003}.
\newblock \bibinfo{title}{{Cosmological smoothed particle hydrodynamics
  simulations: a hybrid multiphase model for star formation}}.
\newblock \bibinfo{journal}{\mnras} \bibinfo{volume}{339},
  \bibinfo{pages}{289--311}.
\newblock \eprint{arXiv:astro-ph/0206393}.
%Type = Article
\bibitem[{{Springel} et~al.(2008){Springel}, {Wang}, {Vogelsberger}, {Ludlow},
  {Jenkins}, {Helmi}, {Navarro}, {Frenk} and {White}}]{2008MNRAS.391.1685S}
\bibinfo{author}{{Springel}, V.}, \bibinfo{author}{{Wang}, J.},
  \bibinfo{author}{{Vogelsberger}, M.}, \bibinfo{author}{{Ludlow}, A.},
  \bibinfo{author}{{Jenkins}, A.}, \bibinfo{author}{{Helmi}, A.},
  \bibinfo{author}{{Navarro}, J.F.}, \bibinfo{author}{{Frenk}, C.S.},
  \bibinfo{author}{{White}, S.D.M.}, \bibinfo{year}{2008}.
\newblock \bibinfo{title}{{The Aquarius Project: the subhaloes of galactic
  haloes}}.
\newblock \bibinfo{journal}{\mnras} \bibinfo{volume}{391},
  \bibinfo{pages}{1685--1711}.
\newblock \eprint{0809.0898}.
%Type = Article
\bibitem[{{Springel} et~al.(2005){Springel}, {White}, {Jenkins}, {Frenk},
  {Yoshida}, {Gao}, {Navarro}, {Thacker}, {Croton}, {Helly}, {Peacock}, {Cole},
  {Thomas}, {Couchman}, {Evrard}, {Colberg} and {Pearce}}]{2005Natur.435..629S}
\bibinfo{author}{{Springel}, V.}, \bibinfo{author}{{White}, S.D.M.},
  \bibinfo{author}{{Jenkins}, A.}, \bibinfo{author}{{Frenk}, C.S.},
  \bibinfo{author}{{Yoshida}, N.}, \bibinfo{author}{{Gao}, L.},
  \bibinfo{author}{{Navarro}, J.}, \bibinfo{author}{{Thacker}, R.},
  \bibinfo{author}{{Croton}, D.}, \bibinfo{author}{{Helly}, J.},
  \bibinfo{author}{{Peacock}, J.A.}, \bibinfo{author}{{Cole}, S.},
  \bibinfo{author}{{Thomas}, P.}, \bibinfo{author}{{Couchman}, H.},
  \bibinfo{author}{{Evrard}, A.}, \bibinfo{author}{{Colberg}, J.},
  \bibinfo{author}{{Pearce}, F.}, \bibinfo{year}{2005}.
\newblock \bibinfo{title}{{Simulations of the formation, evolution and
  clustering of galaxies and quasars}}.
\newblock \bibinfo{journal}{\nat} \bibinfo{volume}{435},
  \bibinfo{pages}{629--636}.
\newblock \eprint{arXiv:astro-ph/0504097}.
%Type = Article
\bibitem[{{Stadel} et~al.(2009){Stadel}, {Potter}, {Moore}, {Diemand}, {Madau},
  {Zemp}, {Kuhlen} and {Quilis}}]{2009MNRAS.398L..21S}
\bibinfo{author}{{Stadel}, J.}, \bibinfo{author}{{Potter}, D.},
  \bibinfo{author}{{Moore}, B.}, \bibinfo{author}{{Diemand}, J.},
  \bibinfo{author}{{Madau}, P.}, \bibinfo{author}{{Zemp}, M.},
  \bibinfo{author}{{Kuhlen}, M.}, \bibinfo{author}{{Quilis}, V.},
  \bibinfo{year}{2009}.
\newblock \bibinfo{title}{{Quantifying the heart of darkness with GHALO - a
  multibillion particle simulation of a galactic halo}}.
\newblock \bibinfo{journal}{\mnras} \bibinfo{volume}{398},
  \bibinfo{pages}{L21--L25}.
\newblock \eprint{0808.2981}.
%Type = Article
\bibitem[{{Stoehr} et~al.(2003){Stoehr}, {White}, {Springel}, {Tormen} and
  {Yoshida}}]{2003MNRAS.345.1313S}
\bibinfo{author}{{Stoehr}, F.}, \bibinfo{author}{{White}, S.D.M.},
  \bibinfo{author}{{Springel}, V.}, \bibinfo{author}{{Tormen}, G.},
  \bibinfo{author}{{Yoshida}, N.}, \bibinfo{year}{2003}.
\newblock \bibinfo{title}{{Dark matter annihilation in the halo of the Milky
  Way}}.
\newblock \bibinfo{journal}{\mnras} \bibinfo{volume}{345},
  \bibinfo{pages}{1313--1322}.
\newblock \eprint{arXiv:astro-ph/0307026}.
%Type = Article
\bibitem[{{Tikhonov} et~al.(2009){Tikhonov}, {Gottl{\"o}ber}, {Yepes} and
  {Hoffman}}]{2009MNRAS.399.1611T}
\bibinfo{author}{{Tikhonov}, A.V.}, \bibinfo{author}{{Gottl{\"o}ber}, S.},
  \bibinfo{author}{{Yepes}, G.}, \bibinfo{author}{{Hoffman}, Y.},
  \bibinfo{year}{2009}.
\newblock \bibinfo{title}{{The sizes of minivoids in the local Universe: an
  argument in favour of a warm dark matter model?}}
\newblock \bibinfo{journal}{\mnras} \bibinfo{volume}{399},
  \bibinfo{pages}{1611--1621}.
%Type = Article
\bibitem[{{Tonry} et~al.(2001){Tonry}, {Dressler}, {Blakeslee}, {Ajhar},
  {Fletcher}, {Luppino}, {Metzger} and {Moore}}]{2001ApJ...546..681T}
\bibinfo{author}{{Tonry}, J.L.}, \bibinfo{author}{{Dressler}, A.},
  \bibinfo{author}{{Blakeslee}, J.P.}, \bibinfo{author}{{Ajhar}, E.A.},
  \bibinfo{author}{{Fletcher}, A.B.}, \bibinfo{author}{{Luppino}, G.A.},
  \bibinfo{author}{{Metzger}, M.R.}, \bibinfo{author}{{Moore}, C.B.},
  \bibinfo{year}{2001}.
\newblock \bibinfo{title}{{The SBF Survey of Galaxy Distances. IV. SBF
  Magnitudes, Colors, and Distances}}.
\newblock \bibinfo{journal}{\apj} \bibinfo{volume}{546},
  \bibinfo{pages}{681--693}.
\newblock \eprint{arXiv:astro-ph/0011223}.
%Type = Article
\bibitem[{{Viel} et~al.(2005){Viel}, {Lesgourgues}, {Haehnelt}, {Matarrese} and
  {Riotto}}]{2005PhRvD..71f3534V}
\bibinfo{author}{{Viel}, M.}, \bibinfo{author}{{Lesgourgues}, J.},
  \bibinfo{author}{{Haehnelt}, M.G.}, \bibinfo{author}{{Matarrese}, S.},
  \bibinfo{author}{{Riotto}, A.}, \bibinfo{year}{2005}.
\newblock \bibinfo{title}{{Constraining warm dark matter candidates including
  sterile neutrinos and light gravitinos with WMAP and the Lyman-{$\alpha$}
  forest}}.
\newblock \bibinfo{journal}{\prd} \bibinfo{volume}{71},
  \bibinfo{pages}{063534}.
\newblock \eprint{arXiv:astro-ph/0501562}.
%Type = Article
\bibitem[{{Wang} and {White}(2007)}]{Wang2007}
\bibinfo{author}{{Wang}, J.}, \bibinfo{author}{{White}, S.D.M.},
  \bibinfo{year}{2007}.
\newblock \bibinfo{title}{{Discreteness effects in simulations of hot/warm dark
  matter}}.
\newblock \bibinfo{journal}{\mnras} \bibinfo{volume}{380},
  \bibinfo{pages}{93--103}.
\newblock \eprint{arXiv:astro-ph/0702575}.
%Type = Article
\bibitem[{{White} and {Rees}(1978)}]{1978MNRAS.183..341W}
\bibinfo{author}{{White}, S.D.M.}, \bibinfo{author}{{Rees}, M.J.},
  \bibinfo{year}{1978}.
\newblock \bibinfo{title}{{Core condensation in heavy halos - A two-stage
  theory for galaxy formation and clustering}}.
\newblock \bibinfo{journal}{\mnras} \bibinfo{volume}{183},
  \bibinfo{pages}{341--358}.
%Type = Article
\bibitem[{{Willick} et~al.(1997){Willick}, {Courteau}, {Faber}, {Burstein},
  {Dekel} and {Strauss}}]{1997ApJS..109..333W}
\bibinfo{author}{{Willick}, J.A.}, \bibinfo{author}{{Courteau}, S.},
  \bibinfo{author}{{Faber}, S.M.}, \bibinfo{author}{{Burstein}, D.},
  \bibinfo{author}{{Dekel}, A.}, \bibinfo{author}{{Strauss}, M.A.},
  \bibinfo{year}{1997}.
\newblock \bibinfo{title}{{Homogeneous Velocity-Distance Data for Peculiar
  Velocity Analysis. III. The Mark III Catalog of Galaxy Peculiar Velocities}}.
\newblock \bibinfo{journal}{\apjs} \bibinfo{volume}{109},
  \bibinfo{pages}{333--+}.
\newblock \eprint{arXiv:astro-ph/9610202}.
%Type = Article
\bibitem[{{Xu} et~al.(2009){Xu}, {Mao}, {Wang}, {Springel}, {Gao}, {White},
  {Frenk}, {Jenkins}, {Li} and {Navarro}}]{2009MNRAS.398.1235X}
\bibinfo{author}{{Xu}, D.D.}, \bibinfo{author}{{Mao}, S.},
  \bibinfo{author}{{Wang}, J.}, \bibinfo{author}{{Springel}, V.},
  \bibinfo{author}{{Gao}, L.}, \bibinfo{author}{{White}, S.D.M.},
  \bibinfo{author}{{Frenk}, C.S.}, \bibinfo{author}{{Jenkins}, A.},
  \bibinfo{author}{{Li}, G.}, \bibinfo{author}{{Navarro}, J.F.},
  \bibinfo{year}{2009}.
\newblock \bibinfo{title}{{Effects of dark matter substructures on
  gravitational lensing: results from the Aquarius simulations}}.
\newblock \bibinfo{journal}{\mnras} \bibinfo{volume}{398},
  \bibinfo{pages}{1235--1253}.
\newblock \eprint{0903.4559}.
%Type = Article
\bibitem[{{Zavala} et~al.(2009){Zavala}, {Jing}, {Faltenbacher}, {Yepes},
  {Hoffman}, {Gottl{\"o}ber} and {Catinella}}]{zavala2009}
\bibinfo{author}{{Zavala}, J.}, \bibinfo{author}{{Jing}, Y.P.},
  \bibinfo{author}{{Faltenbacher}, A.}, \bibinfo{author}{{Yepes}, G.},
  \bibinfo{author}{{Hoffman}, Y.}, \bibinfo{author}{{Gottl{\"o}ber}, S.},
  \bibinfo{author}{{Catinella}, B.}, \bibinfo{year}{2009}.
\newblock \bibinfo{title}{{The Velocity Function in the Local Environment from
  {$\Lambda$}CDM and {$\Lambda$}WDM Constrained Simulations}}.
\newblock \bibinfo{journal}{\apj} \bibinfo{volume}{700},
  \bibinfo{pages}{1779--1793}.
\newblock \eprint{0906.0585}.

\end{thebibliography}
%\end{multicols}
\end{document}